\documentclass{statsoc}

\usepackage{amsmath,natbib}
\usepackage[utf8]{inputenc}
\usepackage{bm}
\usepackage{url}
\usepackage[printonlyused]{acronym}
\usepackage{Sweave}
\usepackage{color}
\usepackage{mathbbol}
\usepackage{numprint}

\renewcommand{\P}{\operatorname{\mathsf{Pr}}} 
\def\eps{\varepsilon}
\def\logit{\text{logit}}
\DeclareMathOperator{\E}{\mathsf{E}} 
\def\SBP{\text{SBP}}
\def\G{\text{G}}

\def\given{\,|\,}
\def\na{\tt{NA}}
\def\nin{\noindent}
\def\mm#1{\ensuremath{\boldsymbol{#1}}} 
\definecolor{grey}{rgb}{.6,.6,.6}
\newcommand{\p}{\operatorname{{p}}} 
\title[Measurement error models using INLA]{Bayesian analysis of measurement error models using INLA}

\author[S. Muff {\it et al.}]{Stefanie Muff$^\star$}
\address{Division of Biostatistics, Institute of Social and Preventive Medicine, University of Zurich, Switzerland.}
\address{Institute of Evolutionary Biology and Environmental Studies, University of Zurich, Switzerland.}
\author{Andrea Riebler$^\star$}
\address{Department of Mathematical Sciences, Norwegian University of Science and Technology, Trondheim, Norway.}
\author{H{\aa}vard Rue}
\address{Department of Mathematical Sciences, Norwegian University of Science and Technology, Trondheim, Norway.}
\author{Philippe Saner}
\address{Institute of Evolutionary Biology and Environmental Studies, University of Zurich, Switzerland.}
\author{Leonhard Held}
\address{Division of Biostatistics, Institute of Social and Preventive Medicine, University of Zurich, Switzerland.}
\coaddress{Leonhard Held, Division of Biostatistics, Institute of Social and Preventive Medicine, University of Zurich, Hirschengraben 84, 8001 Zurich, Switzerland.}
\email{leonhard.held@ifspm.uzh.ch\\ \normalfont $^\star$These authors contributed equally to this work.}

\begin{document}

\begin{abstract}
To account for \ac{ME} in explanatory variables, 
Bayesian approaches provide a flexible framework, as expert knowledge about
unobserved covariates can be incorporated in the prior distributions. 
However, given the analytic 
intractability of the posterior distribution, model inference so far has to 
be performed via time-consuming and complex Markov chain Monte Carlo 
implementations. In this paper we extend the Integrated nested Laplace 
approximations (INLA) approach to formulate Gaussian \ac{ME} models in 
generalized linear mixed models. We present three applications, and show 
how parameter estimates are obtained for common \ac{ME} models, such as the 
classical and Berkson error model including heteroscedastic variances. 
To illustrate the 
practical feasibility, R-code is provided.
\end{abstract}

\acrodef{ME}{Measurement error}

\section{Introduction}

The existence and the effects of measurement error (\ac{ME}) in statistical 
analyses have been recognized and discussed for more than a century, see 
for example \citet{pearson1902, wald1940, berkson1950, Fuller-1987, carroll.etal2006}. 
The sources of \ac{ME} are manifold and imply much more than just instrumental 
imprecision in the measurement of physical variables, such as length, 
weight etc., but may include for instance biases due to preferential sampling, 
incomplete observations or misclassification.

If \ac{ME} is ignored, parameter estimates and confidence intervals in
statistical models often suffer from serious biases. If a regression model is
multivariate and some covariates can be measured with and some without error,
even the effects of the error-free measured covariates can be biased, where the
direction of the bias depends on the correlation among covariates
\citep{carroll.etal1985, gleser.etal1987}.  Moreover, \ac{ME} may cause a loss
of power for detecting signals and connections among variables, and may mask
important features of the data.  Given these facts, it is surprising that
\ac{ME} is often completely ignored or not treated properly. One reason might be
that standard statistical textbooks on regression often pay very little
attention to this aspect, although the problems have been recognized for a long
time.

For successful error-correction both the amount of error 
(i.e.~the error variance) and the error model need to be specified 
correctly. Hence, information about the underlying measurement process is 
essential. Possible errors must be identified early in a study and the entire 
data-collection process should be driven by such considerations.
In the last decades, several approaches to model and correct for \ac{ME} have 
been proposed, such as method-of-moments corrections \citep{Fuller-1987}, 
simulation extrapolation (SIMEX) \citep{cook.stefanski1994}, regression 
calibration \citep{carroll.stefanski1990, gleser1990}, or Bayesian analyses 
\citep{Clayton-1992, stephens.dellaportas1992, richardson.gilks1993, 
dellaportas.stephens1995, gustafson2004}. A thorough overview of current 
state-of-the-art methods is given in the books of \citet{carroll.etal2006} and 
\citet{buonaccorsi2010}. 

In this paper, we focus on Bayesian approaches 
where prior knowledge, and in particular prior uncertainty, 
e.g., in variance estimates, can be incorporated in the model. 
Up to now, posterior marginal distributions in such measurement error 
models have been estimated by employing a  Markov chain Monte Carlo (MCMC) 
sampler, see for example \cite{stephens.dellaportas1992} or 
\cite{richardson.gilks1993}. However, case-specific implementation may be 
challenging, MCMC is time-consuming, and its analysis and interpretation 
requires diagnostic tools.  Generic software like WinBugs 
\citep{lunn-etal-2000}, OpenBugs \citep{lunn.etal2009}, or MCMC samplers in 
{\tt R}, such as MCMCpack \citep{martin-etal-2011}, 
might be used, but they suffer from the same 
drawbacks as any MCMC technique.

Recently, an alternative to MCMC has been proposed to estimate posterior 
marginals by integrated nested Laplace approximations (INLA) for the class 
of latent Gaussian models \citep{rue-etal-2009}. INLA provides accurate 
approximations avoiding time-consuming sampling. Due to its flexibility in 
the choice of likelihood functions and latent models, INLA is an appealing 
alternative to likelihood-based inference in particular for generalized 
linear mixed models (GLMMs) \citep{FongRueWakefield2010}.
The INLA approach is implemented in {\tt C} and easy to use under Linux, 
Windows and Macintosh via a freely available {\tt R}-interface \citep{newR}. 
The {\tt R}-package {\tt r-inla} can be downloaded from \url{www.r-inla.org}. 
Using this package models can be specified in a modular way, where different types 
of regression models can be combined with different types of error models. 
Moreover, it is straightforward to incorporate random effects, such as 
independent or conditional autoregressive (CAR) models to account for 
spatial structure, which is of importance in several settings 
\citep{bernardinelli.etal1997}. Here, we used the {\tt r-inla} version 
updated on July 13, 2013.

In this paper we extend the INLA framework to the most common Gaussian 
\ac{ME} models, namely the classical and the Berkson ME models, which are 
suitable for continuous error-prone covariates. To facilitate the usage of 
the INLA-package with the new features, R-code is provided in the 
Supplementary Material. We hope that the solution presented here will increase 
the use of \ac{ME} thinking in practice and stimulates the greater use of 
Bayesian methods in ME modelling.

Section~\ref{sec:me_examples} introduces three applications from the
biological/medical field containing: a linear regression combined with
heteroscedastic classical error, a logistic model with an binary error-free
covariate and one suffering from classical error, and an overdispersed Poisson
regression model with Berkson error. In Section~\ref{sec:me_models} we will
review the classical and Berkson \ac{ME} models and their effects. Bayesian
analysis with INLA is introduced in Section~\ref{sec:bayes}, where we will
describe how to use this framework for model inference in the presence of
classical and Berkson \ac{ME}.  Section~\ref{sec:applications} presents
modelling details and the results of the three applications analyzed with both
INLA and MCMC. Finally, we provide a discussion and outlook in
Section~\ref{sec:discussion}.

\section{Examples of measurement error problems}\label{sec:me_examples}
In this Section we introduce three applications which will be discussed in more
detail in Section~\ref{sec:applications}. Here, we mainly describe the problem at
hand and the difference of the results depending on whether or not 
measurement error has been incorporated in the analysis. All parameter estimates
in measurement error models are obtained using INLA, as described in detail in
subsequent sections.
\subsection{Inbreeding in Swiss ibex populations}\label{IbexIntro}

We analyzed data described by \cite{bozzuto.etal2013} on $26$ Alpine ibex
populations in Switzerland, some of them monitored over the past 100 years.  The
study aimed to quantify the effect of inbreeding on populations' intrinsic
growth rates. The intrinsic growth rate
$y_i$ of a population $i$ is the theoretical maximal rate of increase, if there
are no density-dependent effects.  The inbreeding coefficient ${x_i}$ of
population $i$ (often denoted as $f_i$) is a quantity between $0$ and $1$, with
larger values indicating stronger inbreeding. Unfortunately, $x_i$ cannot be
measured exactly. A Bayesian analysis based on genotype experiments at $37$
neutral microsatellite loci was employed to derive estimates for $x_i$, denoted
by $w_i$, which additionally provided error variances for each population
$i$. Additional covariates that may influence the intrinsic growth
rate include the number of years a population was observed, the average
precipitation in summer, an interaction between the two, and the average
precipitation in winter.  These covariates are treated as error-free
and subsumed in a row vector $\bm{z}_i$.

Fitting a linear regression model $\E({y}_i)=\beta_0 + \beta_x x_i +
\bm{z}_i \bm{\beta_z}$ in INLA using the proxy $w_i$ instead of the true but
unobserved ${x}_i$, the absolute value of the slope parameter $|\beta_x|$ is
underestimated ($\hat\beta_x=-0.91$, 95\% CI: $[-2.17, 0.36]$). Indeed, after accounting for
\ac{ME} the effect of inbreeding on population growth dynamics is more
pronounced ($\hat\beta_x=-1.84$, 95\% CI: $[-3.88, 0.11]$).

\subsection{Influence of systolic blood pressure on coronary heart disease}\label{FraminghamIntro}

The Framingham heart study is a large cohort study that 
aimed to understand the factors leading to coronary heart 
disease and, in particular, characterize the relation to systolic blood 
pressure~(SBP)~\citep{kannel.etal1986}. The outcome ${y}_i \in \{0,1\}$ is 
a binary indicator for presence of the disease, and modelled via a 
logistic regression. We analysed data from $n=641$ males 
originally presented 
in \cite{macmahon.etal1990}.  As in~\citet[Section 9.10]{carroll.etal2006}, 
we use $x_i=\log(\SBP_i - 50)$ and 
a binary smoking status indicator $z_i \in \{0,1\}$ as predictors. 
The transformation of SBP, originally 
proposed by \cite{cornfield1962}, has also been used in \cite{carroll.etal1984, carroll.etal1996,carroll.etal2006}.
Since it is impossible to measure the long-term SBP, measurements at single 
clinical visits had to be used as a proxy. Note that, due to daily variations 
or deviations in the measurement instrument, the single-visit measures might 
considerably differ from the long-term blood pressure \citep{carroll.etal2006}. 
Hence, the \ac{ME} in SBP has been a concern for many years in this study. 
Importantly, the magnitude of the error could be estimated, as SBP had been 
measured twice at different examinations. These proxy measures for $x_i$ are denoted
as $w_{1i}$ and $w_{2i}$.
A naive approach ignoring \ac{ME} would fit a logistic regression against the 
indicator of coronary heart disease 
$$
\logit\left[\P({y}_i=1 )\right] =\beta_0 +\beta_x {x}_i + \beta_z z_i \ ,
$$ 
where the true covariate ${x}_i$ is replaced by the centered mean of the two
(suitably transformed)
SBP measurements. 
The slope $\beta_x$ is attenuated in this
naive regression ($\hat\beta_x=1.66$, 
95\% CI: $[0.70,2.63]$) compared to the
estimate obtained with error modelling ($\hat\beta_x=1.89$, 95\% CI:
$[0.80,3.00]$).

\subsection{Seedling growth across different light conditions}\label{ShadingIntro}

\begin{figure}
\centering
\mbox{\includegraphics[width=14cm]{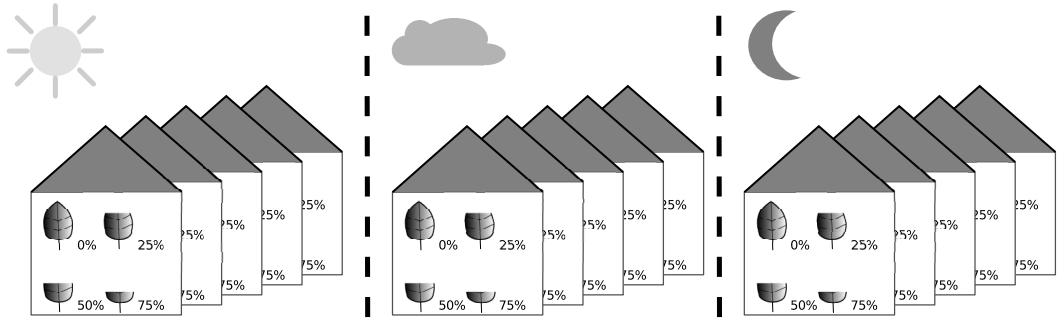}}
\caption{\label{fig:shading}Illustration of the shadehouse experiment. There were five 
shadehouses per light condition and each shadehouse contained four seedlings. 
The seedlings in a shadehouse were each exposed to a different defoliation 
treatment, 0\% indicating that the leaves were not cut, 25\% that one 
fourth of each leaf was cut, etc.}
\end{figure}

The impact of shading (dark, middle, light) and defoliation ($0\%$, $25\%$,
$50\%$, $75\%$ reduction of leaf surface) on plant seedling growth in the Malaysian
rainforest has been investigated in a planned experiment described
in~\cite{paine.etal2012}. The number of new leaves per plant after a four months
growth phase was counted and used as the response variable for plant
growth. Here, we analyzed $60$ seedlings from the species \emph{Shorea fallax},
from which $20$ plants were grown each under dark, middle, and light shading
conditions. There were five shadehouses for each of the three shading
conditions, and each shadehouse contained four seedlings. Each seedling in a
shadehouse was exposed to a different degree of defoliation treatment, compare
Figure~\ref{fig:shading}. In experimental studies in ecology, it is common
practice that the value for the target light intensity $\bm{w}$ (given in \% and
transformed to the log-scale) is assigned
to all replicates within a treatment class (i.e.~dark, middle, light).  However,
due to external conditions the actual observed light availability $\bm{x}$ might
considerably vary from the target value within replicates.  Therefore, the
target light intensity takes only three different values (one for dark, middle
and light), while the actual light availability would take 15 different values
(one for each shadehouse).

The selected regression model is Poisson with (log) target light intensity as
proxy for the actual observed light availability, and additional unstructured
random effects to account for potential overdispersion. 
In contrast to the preceding examples~\ref{IbexIntro} and~\ref{FraminghamIntro}, 
where the inclusion of $\bm{w}$ instead of $\bm{x}$ in the regression 
attenuates the parameter estimates, theory for 
log-linear models with Berkson error suggests that there is
no bias in the regression coefficients~\citep{carroll.1989}. However, it is not clear if this result
extends to models with random effects. Our analysis did not
reveal a difference in the regression coefficients after accounting for measurement
error. We did observe a slightly increased credible interval width for the
regression coefficients, and, in particular, for the precision of the random effects.

\section{Measurement error models in regression}\label{sec:me_models}
\subsection{The generalized linear model}

Assume we have $n$ observations in a generalized linear model (GLM). 
The data are given as $(\bm{y},\bm{z},\bm{x})$, with 
$\bm{y}=(y_{1},\ldots,y_n)^\top$ denoting the response, 
$\bm{z}=(\bm{z}_1,\ldots,\bm{z}_p)$ a covariate matrix of 
dimension $n \times p$ for $p$ error-free covariates, and 
$\bm{x}=({x}_1,\ldots,{x}_n)^\top$ a single error-prone covariate 
whose true values are unobservable. The generalization to multiple 
error-prone covariates is straightforward. Suppose $\bm{y}$ is of 
exponential family form with mean $\mu_{i}= \E(y_{i}\given \bm{x}, \bm{z}, \bm{\beta})$, 
linked to the linear predictor $\eta_{i}$ via
\begin{align}
\mu_{i} &= h(\eta_{i}) \nonumber \\
 \eta_{i} &= \beta_0 + \beta_x x_{i} + \bm{z}_{[i,]}\bm\beta_z \ .\label{eq:glmb}
\end{align}
Here, $h(\cdot)$ is a known monotonic inverse link (or response) function, 
$\beta_0$ denotes the intercept, $\beta_x$ the fixed effect for the 
error-prone covariate $\bm{x}$, and $\bm{z}_{[i,]}$ is $1 \times p$ with a 
corresponding vector $\bm\beta_z$ of fixed effects.
This GLM is extended to a generalized linear mixed model (GLMM) by adding 
normally distributed random effects on the linear predictor scale (\ref{eq:glmb}). 

\begin{figure}
\centering
\mbox{\includegraphics[width=6cm]{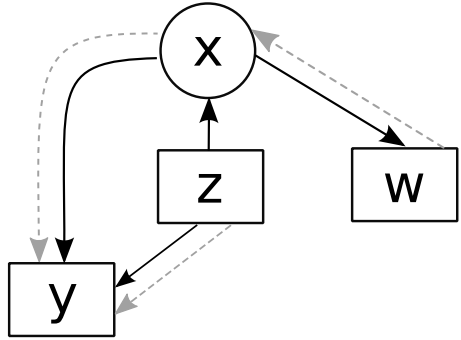}}
\caption{\label{fig:graph_mod}Graphical structure of the error models. Rectangular boxes 
illustrate variables that are observed, while circles indicate 
unknown variables. Black arrows correspond to the classical, and the 
dashed grey arrows to the Berkson error model. Note the change in direction 
of the arrow between $\bm{x}$ and $\bm{w}$.}
\end{figure}

Let $\bm{w} = (w_1, \ldots, w_n)^\top$ denote the observed version of the 
true, but unobserved covariate $\bm{x}$. We distinguish two 
different \ac{ME} processes: the classical error model and the Berkson 
error model \citep{berkson1950}. The graphical structure of these models 
is very similar, compare Figure~\ref{fig:graph_mod}, but the caused effects 
are fundamentally different.

\subsection{Classical measurement error model}\label{sec:classical}
In the classical error model it is assumed that the covariate $\bm{x}$ 
can be observed only via a proxy $\bm{w}$, such that, in vector notation,
\begin{equation*}
\bm{w} = \bm{x} + \bm{u} \ ,
\end{equation*}
with $\bm{u}=(u_1, \ldots, u_n)^\top$.
Throughout the paper the components of 
the error vector $\bm{u}$ are assumed to be independent and normally 
distributed with mean zero and variance $\tau_u^{-1}$, 
i.e.~$\text{Cov}(u_i, u_j) = 0$ for $i \ne j$. Note that in the following 
we parameterize the normal distribution with mean and precision 
(or precision matrix in the multivariate context), rather than using 
the variance or covariance matrix. 

We assume that the error term $\bm{u}$ 
is independent of the true covariate $\bm{x}$, but also independent 
of any other covariates $\bm{z}$ and the response $\bm{y}$. This implies 
a non-differential \ac{ME} model, meaning that $\bm{y}$ and $\bm{w}$ are 
conditionally independent given $\bm{z}$ and $\bm{x}$. In most 
applications this assumption is plausible as it implies that, given the 
true covariate $\bm{x}$ and covariates $\bm{z}$, no additional information 
about the response variable $\bm{y}$ is gained through $\bm{w}$ 
\citep{carroll.etal2006}.
Ideally, repeated measurements $w_{ji}$, $j=1, \ldots, J$, of the true value $x_{i}$ are available, so that
\begin{equation}
 {w_{ji}} \given  {x_{i}} \sim \mathcal{N}( {x_{i}},\tau_{u}) \ .\label{ClErrorModel}
\end{equation}
More generally, the error structure can be heteroscedastic with $\bm{w}_{j}\sim\mathcal{N}(\bm{x},\tau_u\mathbf{D})$, where $\bm{w}_{ j}$ denotes the vector of the $j^{th}$ measurements, 
and the entries in the 
diagonal matrix $\mathbf{D}$ represent weights $d_i$ that are proportional 
to the individual error precision $\tau_u(x_i)$ depending on $x_i$, 
which allows for a heteroscedastic error structure. This is required when the 
accuracy of surrogate $w_i$  depends on $i$, i.e.,~$x_i$ can be measured
with varying accuracy for different $i$.
In fact, both the homo- and heteroscedastic cases are relevant in practice 
(see, e.g., \cite{subar.etal2001} or example \ref{ibex} presented here).

\begin{figure}
\centering
\mbox{
\setkeys{Gin}{width=\textwidth}
\includegraphics{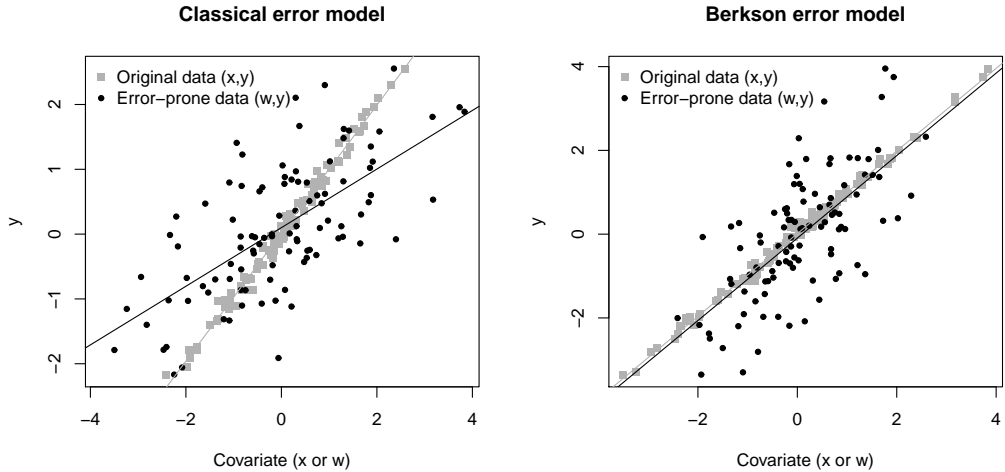}
}
\caption{\label{fig:me-illustration}Effect of \ac{ME} in the linear model. Left: Classical \ac{ME}. 
Two effects can be seen: 1) The absolute value of the covariate estimate 
is biased (attenuated); 2) The variability around the regression line in 
the data with \ac{ME} (black circles) is much larger than in the case of 
the truly observed data (grey squares). Right: Berkson \ac{ME}. The absolute 
value of the covariate estimate is unbiased in the linear model, while the 
variability around the regression line is larger for the data with \ac{ME}.} 
\end{figure}

Estimates of $\beta_x$ are usually attenuated in the classical \ac{ME} setting
if $\bm{w}$ is taken as a proxy for $\bm{x}$.  Consider for instance a simple
linear regression with homoscedastic \ac{ME}. Fitting the {\em naive} model
$\bm{y}=\beta_0^\star\mm{1}+\beta_x^\star\bm{w}+\bm{\eps}^\star$ instead of the true
model $\bm{y}=\beta_0\mm{1}+\beta_x\bm{x}+\bm{\eps}$ will result in
$|\beta_x^\star|<|\beta_x|$, if the error variance $1/\tau_u$ is larger than
zero. 
The left panel of
Figure~\ref{fig:me-illustration} illustrates this attenuation affect. Another
important effect is the significant increase of the variability around the
regression line.

\subsection{Berkson measurement error model}
Berkson-type error can be observed in experimental settings, where the 
value of a covariate may correspond to, e.g.,~a predefined fixed dose, 
temperature or time interval, but the true values $\bm{x}$ may deviate 
from these planned values $\bm{w}$ due to imprecision in the realization. 
The second setting where Berkson-type error occurs is in epidemiological 
or biological studies, where, e.g., averages of exposures in areas are 
assigned to individuals living or working close-by.
Examples are the application of fixed doses of herbicides in bioassay 
experiments \citep{rudemo.etal1989} or the radiation epidemiology study 
described in \cite{kerber.etal1993} and \cite{simon.etal1995}.
Such circumstances led to the Berkson error model \citep{berkson1950}
\begin{equation*}
\bm{x} = \bm{w} + \bm{u} \ ,
\end{equation*}
where $\bm{u}$ and $\bm{w}$ are independent, and 
\begin{equation}
 \bm{x} \, | \,  \bm{w} \sim \mathcal{N}( \bm{w},\tau_u\mathbf{D}) \ , \label{BErrorModel}
\end{equation}
with $\mathbf{D}$ denoting a diagonal matrix as in Section~\ref{sec:classical}. 
Like classical \ac{ME}, the Berkson error is also assumed to be non-differential. 
The effect of Berkson error is fundamentally different from that of classical 
error. In the linear regression model there is no attenuation effect, as 
illustrated in the right panel of Figure~\ref{fig:me-illustration}. However, 
the residual precision suffers from the same qualitative bias as in the 
classical \ac{ME} model. Issues become more involved for GL(M)Ms. 
For instance, parameter estimates for logistic regression are only 
approximately consistent in the Berkson case \citep{Burr1988, bateson.etal2010}, 
which makes error modelling essential.

The difference between the classical and the Berkson error model
is reflected in the relationships between the error variances. 
Denote with $\tau_x^{-1}$ and $\tau_{w}^{-1}$ the variances of $\bm{x}$ 
and $\bm{w}$, respectively. Due to the independence assumption of $\bm{x}$ 
and $\bm{u}$ in the classical and between
$\bm{w}$ and $\bm{u}$ in the Berkson error case, the variances in the two 
\ac{ME} models can be written as
\begin{eqnarray*}
\tau_w^{-1} &=& \tau_x^{-1} + \tau_u^{-1} \quad \text{(classical)}\ ,\\
\tau_x^{-1} &=& \tau_w^{-1} + \tau_u^{-1} \quad \text{(Berkson)} \ .
\end{eqnarray*}
Thus, the surrogate $\bm{w}$ is more variable than the true 
covariate $\bm{x}$ in the classical model, whereas the opposite is true 
in the Berkson case. This effect can also be observed in Figure~\ref{fig:me-illustration}.

\section{Analysis of measurement error models using INLA}\label{sec:bayes}
A Bayesian analysis of \ac{ME} models dates back to the seminal work of
\citet{Clayton-1992} and is based on a three-level hierarchical model.  The
first level represents the observation model $\bm{y}\,|\,\bm{v}, \bm{\theta}_1$
defining distributional assumptions about the response variable $\bm{y}$ in
dependence on some unknown (latent) parameters $\bm{v}$ and certain
hyperparameters $\bm{\theta}_1$, e.g.~variance or correlation parameters. The
second level describes the latent model or unobserved process
$\bm{v}\,|\,\bm{\theta}_2$ depending on hyperparameters $\bm{\theta}_2$. In the
third level, hyperpriors are defined for the hyperparameters $\bm{\theta} =
(\bm{\theta}_1^\top, \bm{\theta}^\top_2)^\top$.  The posterior distribution of
$\bm{v}$ and $\bm{\theta}$ is then given by
\begin{equation}\label{eq:posterior}
  \p(\bm{v}, \bm{\theta}\,|\, \bm{y}) \propto \p(\bm{y}\,|\,\bm{v}, \bm{\theta}) \p(\bm{v}\,|\,\bm{\theta}) \p(\bm{\theta}).
\end{equation}

Of primary interest are often the posterior marginal distributions of 
components $v_i$ of $\bm{v}$, which can be derived from~\eqref{eq:posterior} via
\begin{align}
  \p(v_{i}\given\bm{y}) &= \int_{\bm{\theta}}\int_{\bm{v}_{-i}}\p(\bm{v}, \bm{\theta}\,|\, \bm{y})d\bm{v}_{-i}d\bm{\theta}\ , \label{eq:marginal_general}
\end{align}
as well as posterior marginals of the hyperparameters $\theta_j$.  The
computation of massively high integrals is however very difficult.  Except for
cases where everything can be computed analytically, exact inference is 
challenging. Hence, sampling-based approaches have been the
standard tool \citep{GelfandSmith1990}.  Currently, few generic software
packages based on MCMC, e.g.~OpenBugs \citep{lunn.etal2009}, are available.
However, MCMC based approaches are time-consuming and require diagnostic checks
to ensure good mixing properties and convergence of the simulated samples.

\cite{rue-etal-2009} proposed with INLA an efficient computing methodology 
based on accurate approximations to perform Bayesian inference in a 
sub-class of hierarchical models, namely latent Gaussian models. In this class
the second level, the latent model, is assumed to be Gaussian. 
In the following we will shortly present the general idea of INLA.

INLA uses the fact that Equation \eqref{eq:marginal_general} can also be written as
\begin{align*}
  \p(v_{i}\given\bm{y}) &= \int_{\bm{\theta}}\p(v_{i}\given\bm{\theta}, \bm{y})\p(\bm{\theta}\given\bm{y})d\bm{\theta}\ ,
\end{align*}
and approximates this term by the finite sum
\begin{align*}
  \tilde{\p}(v_{i}\given\bm{y}) &= \sum_k \tilde{\p}(v_{i}\given\bm{\theta}_k, \bm{y})\tilde{\p}(\bm{\theta}_k\given\bm{y}) \Delta_k\ . 
\end{align*}
Here, $\tilde{\p}(v_{i}\given\bm{\theta}, \bm{y})$ and $\tilde{\p}(\bm{\theta}\given\bm{y})$
denote approximations of  $\p(v_{i}\given\bm{\theta}, \bm{y})$
and $\p(\bm{\theta}\given\bm{y})$, respectively. For ${\p}(\bm{\theta}\given\bm{y})$ a Laplace
approximation is used, while for $\p(v_{i}\given\bm{\theta}, \bm{y})$ three
different strategies are available, see \cite{rue-etal-2009}. The default is a 
simplified Laplace approximation. Finally, the sum is computed over suitable support
points $\theta_k$ with appropriate weights $\Delta_k$. Posterior marginals for
$\p(\theta_j \given \bm{y})$ can be obtained similarly from $\tilde{\p}(\bm{\theta}\given\bm{y})$.
INLA can be used via the {\tt R}-package {\tt r-inla}, and is called in a modular way. Different types
of likelihood functions in the first level can thus be combined with different regression models
in the second level.
As discussed in \cite{rue-etal-2009} and illustrated in a variety of different 
applications, the approximation error of INLA is small compared to the Monte 
Carlo error and is negligible in practice, see for example 
\cite{paul-etal-2010, schroedle.etal2011, riebler-etal-aoas-2012}. 

\subsection{Classical measurement error - general case}\label{sec:classicalINLA}

In the following we show how the classical measurement error model fits
into the hierarchical structure required by INLA. 
Consider a generalized linear (mixed) model regressing a response
variable $\bm{y}$ on covariates $\bm{x}$ and $\bm{z}$. 
The $p$ covariates in $\bm{z}$ can be observed directly, while instead of
$\bm{x}$ only a surrogate $\bm{w} | \bm{x},\bm{\theta} \sim \mathcal{N}(\bm{x}, \tau_u \mathbf{D})$,
following the classical error model (\ref{ClErrorModel}), is available.
The distribution of $\bm{x}$, possibly depending on $\bm{z}$, is specified
in the \emph{exposure model}~\citep{gustafson2004}.
%
%
In the most general case considered here, the covariate $\bm{x}$ is Gaussian with mean depending on
$\bm{z}$, i.e. 
\begin{equation}
\bm{x} \given \bm{z} \sim \mathcal{N}(\alpha_0\mm{1}+ \bm{z}\bm{\alpha}_z, \tau_{x}\mathbf{I}) \ . \label{eq:ExpModel}
\end{equation}
Here, $\alpha_0$ is the intercept, $\bm{\alpha}_z$ 
  is
the $p\times 1$ vector of fixed effects, and $\tau_{x}^{-1}$ the residual
variance in the linear regression of $\bm{x}$ on
$\bm{z}$. 
If $\bm{x}$
depends only on certain components of $\bm{z}$, then the corresponding entries
in $\bm{\alpha}_z$ are set to zero. The extreme case $\bm{\alpha}_z=\bm{0}$,
where $\bm{x}$ is independent of $\bm{z}$, is discussed separately in
Section \ref{sec:mec.model}.

The assumption that the distribution of the unobserved $\bm{x}$ given the 
observable covariates~$\bm{z}$ follows a normal distribution is crucial to apply
INLA, but often justified. Due to recent extensions of INLA, see
\citet{martins.rue2012}, $\bm{x} | \bm{z}$ could even follow a non-Gaussian
distribution, so that the normal assumption might be relaxed in the future.

The general model structure including \eqref{eq:glmb},
\eqref{ClErrorModel} and \eqref{eq:ExpModel} is hierarchical.
The first level in the hierarchical Bayesian analysis encompasses 
three models, i.e., the regression model, exposure model and error model
\begin{eqnarray}
\E(\bm{y}) &= & h(\beta_0\mm{1} + \beta_x \bm{x} +  \bm{z}\bm\beta_z) \ , \label{eq:regMod1}\\
\bm{0}  &=& -\bm{x} + \alpha_0\mm{1} + \bm{z} \bm\alpha_z + \bm\eps_{x} \ , \quad \bm\eps_{x} \sim  \mathcal{N}(0,\tau_{x}\mathbf{I}) \ , \label{eq:inlaExp} \\
\mbox{ and }
\bm{w}  &=& \bm{x} + \bm{u} \ , \qquad  \;\; \bm{u} \sim  \mathcal{N}(0,\tau_u\mathbf{D}) \ . \label{eq:inlaError} 
\end{eqnarray}
Of note, in \eqref{eq:inlaError} $\bm{w}$ represents the stacked vector of the 
repeated measurements $\bm{w}_1, \ldots, \bm{w}_J$, $\bm{x}$ is 
repeated accordingly $J$ times, so that it has the same length, 
and $\mathbf{D}$ is of appropriate dimension.
Implementation in INLA requires a \emph{joint model} formulation,
where the response variable $\bm{y}$ is augmented with 
pseudo-observations $\bm{0}$, compare equation \eqref{eq:inlaExp}, 
and the observed values $\bm{w}$ of the measurement error model
\eqref{eq:inlaError}. 
Note that the exposure model \eqref{eq:ExpModel}, encoded in \eqref{eq:inlaExp},
can be easily extended to include structured or unstructured random effects terms.
The resulting response matrix for {\tt r-inla} contains one separate 
column per equation, namely
$$
\begin{bmatrix}
y_{1} & \na & \na \\
\vdots & \vdots & \vdots \\
y_{n} & \na & \na \\
\na & 0 & \na\\
\vdots & \vdots & \vdots \\
\na & 0 & \na\\
\na & \na & {w}_{11} \\
\vdots & \vdots & \vdots \\
\na & \na & {w}_{1n} \\
\na & \na & w_{21} \\
\vdots & \vdots & \vdots \\
\na & \na & w_{(J-1)n} \\
\na & \na & {w}_{J1} \\
\vdots & \vdots & \vdots \\
\na & \na & {w}_{Jn} \\
\end{bmatrix} \ .
$$ 
Each column requires specification of a likelihood function.  The
first follows the selected exponential family distribution for the
response $\bm{y}$ with mean
\eqref{eq:regMod1}. The second is assumed to be Gaussian, see
\eqref{eq:inlaExp}. The third component is also Gaussian, 
as specified in \eqref{eq:inlaError}.

The second level of this hierarchical model is formed by the latent field
$\bm{v}=(\beta_0, \bm{\beta}_z^\top, \alpha_0, \bm{\alpha}_z^\top,
\bm{x})^\top$.  Note that the regression coefficient $\beta_x$ is not included
in the latent field $\bm{v}$, but is considered as an unknown 
hyperparameter, so $\bm{\theta}=(\beta_x, \tau_u, \tau_x,
\bm{\theta}_1^\top)^\top$. Here, $\bm{\theta}_1$ represents additional
hyperparameters of the observation model~\eqref{eq:regMod1}.

Gaussian prior distributions are now assigned to the components of $\bm{v}$.  We
use independent normal prior distributions with zero mean and small precision
for $\beta_0$ and the components of $\bm{\beta_z}$.  Further, we try to elicit
mean and precision of $\alpha_0$ and $\bm{\alpha_z}$ by incorporating
prior/expert knowledge about the distribution of $\bm{x} | \bm{z}$. In the
simplest case, the components of $\bm{x}$ are assumed to be independent and
identically distributed, but this can be relaxed, if appropriate.

Importantly, the latent component $\bm{x}$ appears in all three observation
models \eqref{eq:regMod1}, \eqref{eq:inlaExp}, and \eqref{eq:inlaError}.
To integrate the product $\beta_x \bm{x}$ from~\eqref{eq:regMod1} in the
formulation, an almost identical copy $\bm{x}^\star$ of $\beta_x \bm{x}$ is
created.  This is achieved by extending the latent
vector $\bm{x}$ to $\bm{x}_c = (\bm{x}^\top, {\bm{x}^\star}^\top)^\top$ with $\pi(\bm{x}_c) = \p(\bm{x})
\p(\bm{x}^\star \given \bm{x})$ and
\begin{equation*}
	\p(\bm{x}^\star\given\bm{x}, \beta_x, \tau) \propto \exp\left(-\frac{\tau}{2} (\bm{x}^\star - \beta_x \bm{x})^\top(\bm{x}^\star - \beta_x \bm{x})\right).
\end{equation*}
The precision $\tau$, fixed to some large value, controls the similarity between
$\bm{x}^\star$ and $\beta_x \bm{x}$ (default value: $10^9$). The regression 
coefficient $\beta_x$ is treated as unknown, in contrast to other
applications of the copy function in INLA, where the respective coefficient 
is often equal to one
\citep{martins.etal2012}.  For exact specification within {\tt r-inla},
compare application~\ref{Framingham} and the corresponding Supplementary
Material.

The third level concerns the hyperpriors. In our applications we assume a
normal distribution with mean zero and low precision for $\beta_x$. For $\tau_x$
and $\tau_u$ we assume gamma distributions where the corresponding shape and
scale parameters are chosen based on expert knowledge. Other prior distributions
for $\tau_x$ and $\tau_u$ can be used in INLA, see \citet{roos.held2011} for
an example. Even user-defined (non-standard) priors, which can be specified
using a grid of $x$- and $y$-values, are supported.

\subsection{Classical measurement error - independent exposure model}\label{sec:mec.model}

To facilitate the integration of simple \ac{ME} models in INLA, we also provide
a specific \ac{ME} model called {\tt mec} within the {\tt r-inla} software,
which does not require specification of a joint model using the copy function.
The tool covers the case where the exposure model for $\bm{x}$ is independent of the other covariates $\bm{z}$, i.e., the general exposure model~\eqref{eq:ExpModel} reduces to 
\begin{equation*}
\bm{x} \sim \mathcal{N}(\alpha_0\mm{1}, \tau_x \mathbf{I}) \ .
\end{equation*} 
Its derivation is
sketched in the following and its use is shown in Section~\ref{ibex} and the
corresponding Supplementary Material.

Without loss of generality, we can omit the parameters $\beta_0$ and
$\bm{\beta}_z$ in \eqref{eq:regMod1} and consider the simplified model
\begin{align}
\E(\bm{y}) &= h(\beta_x   \bm{x})  \ , \label{eq:eta}\\
 \bm{x}  &=  \alpha_0\mm{1} + \bm{\epsilon}_x \ , \quad \bm\eps_{x} \sim
 \mathcal{N}(0,\tau_{x}\mathbf{I}) \ , \nonumber \\
\mbox{ and } \bm{w}  &=   \bm{x}   + \bm{u} \ , \qquad  \;\; \bm{u} \sim  \mathcal{N}(0,\tau_u\mathbf{D}) \ . \nonumber 
\end{align}
Here, $\alpha_0$ is also considered a hyperparameter, thus the latent field $\bm{v}$ now only contains $\bm{x}$, leading to $\bm{\theta} = (\beta_x, \tau_x,
\tau_u, \alpha_0)^\top$. The posterior distribution of $\bm{x}$ and
$\bm{\theta}$ is then
\begin{eqnarray*}
    \p(\bm{x}, \bm{\theta} \given \bm{y}, \bm{w}) & \propto & \p(\bm{\theta})
    \; \p(\bm{x} \given\bm{\theta}) \; \p(\bm{w} \given \bm{x}, \bm{\theta}) \;
    \p(\bm{y} \given \bm{x}, \bm{\theta}) \\
    & \propto & \p(\bm{\theta})\;
    \p(\bm{x} \given \bm{w}, \bm{\theta}) \; \p(\bm{w} \given \bm{\theta}) \;
    \p(\bm{y} \given \bm{x}, \bm{\theta}) \ ,\label{eq:modlik}
\end{eqnarray*}
using $ \p(\bm{x} \given \bm{\theta})  \p(\bm{w} \given \bm{x}, \bm{\theta}) = \p(\bm{x} \given \bm{w}, \bm{\theta})  \p(\bm{w} \given \bm{\theta})$.
Now
\begin{equation}
    \bm{w} \given \bm{\theta} \;\sim\; {\mathcal N}\left(
      \alpha_0 \mm{1}, \left[(\tau_u\mathbf{D})^{-1} + (\tau_x\mathbf{I})^{-1}\right]^{-1} \right)\ \nonumber
\end{equation}
and
\begin{eqnarray*}
\p(\bm{x} \given \bm{w}, \bm{\theta}) &\propto & \p(\bm{x} \given \bm{\theta})\p(\bm{w}\given \bm{x},\bm{\theta}) \\
 & \propto & \exp\left(-\frac{\tau_x}{2}(\bm{x}-\alpha_0\mm{1})^\top (\bm{x}-\alpha_0\mm{1}) -\frac{\tau_u}{2} (\bm{x}-\bm{w})^\top \mathbf{D} (\bm{x}-\bm{w})  \right) \ .
\end{eqnarray*}
Combining these quadratic forms gives
\begin{equation}
    \bm{x} \given \bm{w}, \bm{\theta} \;\sim\;{\mathcal N}\left(
      (\tau_{x} \alpha_0 \mm{1} + \tau_u \mathbf{D} \bm{w} ) ( {\tau_{x}} \mathbf{I} + \tau_u \mathbf{D})^{-1}\ , \,
       {\tau_{x}} \mathbf{I} +  \tau_u \mathbf{D} \right), \nonumber
\end{equation}
so the posterior distribution $\p(\bm{x}, \bm{\theta} \given \bm{y}, \bm{w})$
can be evaluated explicitly. An alternative formulation can be
obtained by considering
$\bm{\nu} = \beta_x\bm{x}$ instead of $\bm{x}$, where
\begin{equation*}
	\bm{\nu} \given \bm{w}, \bm{\theta} \;\sim\;{\mathcal N}\left(
      \beta_x (\tau_{x} \alpha_0 \mm{1} + \tau_u \mathbf{D} \bm{w} ) ( {\tau_{x}} \mathbf{I} + \tau_u \mathbf{D})^{-1}, \,
       \frac{{\tau_{x}} \mathbf{I} +  \tau_u \mathbf{D}}{\beta_x^2} \right) \ . \nonumber
\end{equation*}
This model is termed {\tt mec} within the {\tt r-inla} and has four
hyperparameters: $\beta_x$, $\tau_x$, $\tau_u$, and $\alpha_0$. Its advantage 
is a considerable simplification of the {\tt r-inla} call, see the Supplementary Material for
code examples.

\subsection{Berkson measurement error}

We again consider a generalized linear (mixed) model \eqref{eq:glmb}, but
replace the classical error model \eqref{ClErrorModel} by the Berkson model
\eqref{BErrorModel}
\begin{eqnarray*}
\bm{x}\given\bm{w},\bm{\theta} \sim \mathcal{N}(\bm{w},\tau_u\mathbf{D}) \ .
\end{eqnarray*}
Since $\bm{x}$ is defined conditionally on the observations $\bm{w}$, the
exposure model \eqref{eq:ExpModel} becomes obsolete. Analogous to Section
\ref{sec:mec.model}, where $\bm{x}$ did not depend on the other covariates
$\bm{z}$, we can define a latent Gaussian model for the Berkson measurement
error model. Indeed, the same simplifications as in \eqref{eq:eta} lead to the
hierarchical model
\begin{align}
\E(\bm{y}) &=  \beta_x   \bm{x}   \ , \nonumber\\
  \bm{x}  &= \bm{w}   + \bm{u} \ , \qquad  \;\; \bm{u} \sim  \mathcal{N}(0,\tau_u\mathbf{D}) \ \nonumber \ ,
\end{align}
where $\bm{x}$ is the latent field and 
the hyperparameters are $\bm{\theta} = (\beta_x,\tau_u)^\top$. 
Importantly, the latent model $\bm{x}|\bm{w},\bm{\theta}$ is now identical 
to the error model (\ref{BErrorModel}), because the latent field $\bm{v}$ 
contains only $\bm{x}$. It is thus straightforward to calculate the posterior 
distribution
\begin{equation*}
    \p(\bm{x}, \bm{\theta} \given \bm{y}, \bm{w}) \;\propto\; \p(\bm{\theta})
    \; \p(\bm{x} \given \bm{w},\bm{\theta}) \;
    \p(\bm{y} \given \bm{x}, \bm{\theta}) \ .
\end{equation*}

The reparameterization $\bm{\nu} = \beta_x\bm{x}$ is again useful and leads to
\begin{equation*}
\bm{\nu} \given \bm{w},\bm{\theta} \;\sim\;{\mathcal N}\left(
      \beta_x \bm{w}, \,
     \frac{\tau_u}{\beta_x^2}\mathbf{D} \right) \ .
\end{equation*}
This model is termed ``{\tt meb}'' within the {\tt R}-package {\tt r-inla} 
and has two hyperparameters: $\beta_x$ and $\tau_u$.

As in Section~\ref{sec:classicalINLA}, the copy function can also be used
for Berkson measurement error models. 
However, here it does not add to the generality of the model specification 
as no exposure model is involved in Berkson measurement error models. Thus, 
we recommend the use of the {\tt meb} model, and just 
illustrate for completeness the formulation using the copy function.
Since the respective joint model contains only the two components
\begin{eqnarray*}
\E(\bm{y}) &=& h(\beta_0\mm{1} + \beta_x \bm{x} +  \bm{z}\bm\beta_z) \ , \label{regMod2}\\
-\bm{w}&=&-\bm{x} + \bm{u} \ , \label{BerksonMod}
\end{eqnarray*}
the response matrix simplifies to
$$
\begin{bmatrix}
y_{1} & \na  \\
\vdots & \vdots \\
y_{n} & \na \\
\na &  -w_1 \\
\vdots &   \vdots \\
\na &  -w_n \\
\end{bmatrix} \ .
$$
The latent field is now given by
$\bm{v}=(\bm{x}^\top,\beta_0,\bm\beta^\top_z)^\top$ 
and $\bm{\theta}=(\beta_x,\tau_u, \bm\theta_1^\top)^\top$ are the hyperparameters,
where $\bm\theta_1$ may again contain additional hyperparameters of the
likelihood. As before, all components in $\bm{v}$ and the coefficient $\beta_x$ obtain
Gaussian priors, and the error precision $\tau_u$ a suitable gamma prior.

\section{Applications}\label{sec:applications}

In the following we demonstrate how to define the different measurement 
error applications introduced in Section~\ref{sec:me_examples} in the INLA 
framework. The respective {\tt r-inla} code is given in the Supplementary 
Material. A comparison of the results obtained by INLA to those obtained by 
an independent MCMC implementation is provided for each application to 
highlight the accuracy of INLA. The
efficiency of MCMC might be reduced when using uncentered covariates \citep{gelfand.etal1995, gelfand.etal1996}, 
and additional adjustments of the default parameters in
the numerical optimization routine of INLA might be needed. Hence,
we center all continuous covariates around zero in the following analyses.

\subsection{Inbreeding in Swiss ibex populations} \label{ibex}
\begin{figure}
\centering
\mbox{\includegraphics[width=7cm]{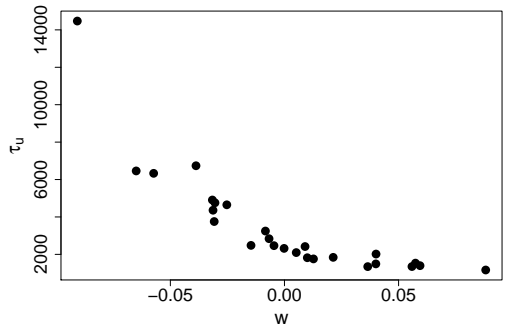}}
\caption{\label{IbexError}Uncertainty in the covariate $\bm{x}$ in the 
ibex study, depending on the estimate $\bm{w}$. Larger values can be 
estimated with less precision (i.e.~larger variance 1/$\tau_{u}$).}
\end{figure}

The ibex data introduced in Section~\ref{IbexIntro} were analyzed using 
a linear model with classical heteroscedastic error variances.
The observation model is a Gaussian 
\begin{equation*}
\bm{y} \,|\, \bm{x}  \sim \mathcal{N}(\beta_0\mm{1}+\beta_x\bm{x} + \bm{z} \bm{\beta}_z,\tau_\eps\mathbf{I}) 
\end{equation*}
with $\bm{y}$ being the intrinsic growth rates, $\bm{x}$ the inbreeding 
coefficients of the populations, and $\bm{z}$ the matrix of additional
covariates, as listed in Section~\ref{IbexIntro}.
The level of inbreeding $x_i$ in population $i=1,\ldots, 26$ was estimated 
as $w_i$ from a Bayesian analysis, which, as a by-product, also provided 
an estimated population-specific error precision $\hat\tau_{u_i}$. 
Since larger values of $\bm{w}$ have more uncertainty, i.e.~smaller precision, 
as shown in Figure~\ref{IbexError}, it is natural to formulate a 
heteroscedastic classical error model
\begin{equation*}
\bm{w} \, | \,  \bm{x} \sim \mathcal{N}( \bm{x},\tau_{u}\mathbf{D}) 
\end{equation*}
with entries $\hat\tau_{u_i}$ in the diagonal matrix $\mathbf{D}$. 
Since $\bm{x}$ is assumed to be uncorrelated to the covariates $\bm{z}$, 
the exposure model~\eqref{eq:ExpModel} reduces to
\begin{equation*}
\bm{x} \sim \mathcal{N}(\alpha_0\mm{1},\tau_x\mathbf{I}) \ .
\end{equation*}
Here, $\alpha_0=0$ was fixed due to the preceding covariate centering. 
The unknowns in this example are the latent field 
$\bm{v}=(\bm{x}^\top,\beta_0, \bm\beta^\top_z)^\top$ and the 
hyperparameters $\bm\theta=(\beta_x,\tau_u,\tau_x,\tau_\eps)^\top$.
We assigned independent $\mathcal{N}(0, 10^{-4})$ priors to all
$\beta$-coefficients.  The assignment of the prior distributions to the
precision parameters is more delicate. We used gamma distributions, where the
corresponding shape and rate parameters were chosen based on expert/prior
knowledge. In practice the inbreeding coefficient $\bm{x}$ of sexually breeding
species is not observed over the whole theoretical range $[0,1]$. For
populations of similar age and size as in the current study, $\bm{x}$ values
are expected to lie within $[0, 0.45]$~\citep{biebach.keller2010}.  Assuming a
uniform distribution within this range, this corresponds to the precision
$12/0.45^2 \approx 59$, which we take as a lower limit for $\tau_x$.
In the absence of prior knowledge from other 
studies, we assume that the range of $\bm{x}$ is at least $0.05$, which gives 
an upper limit of $4800$, again assuming a uniform distribution. 
A gamma distribution with $2.5\%$ quantile at $59$ and $97.5\%$ quantile 
at $4800$ is determined by numerical optimization, resulting in $\G(1, 0.0009)$.

The precision $\tau_u$ represents a possible multiplicative bias in the
estimates $\hat{\tau}_{u_i}$. Here, we assume that this bias is uniformly
between $0.5$ and $2$ with probability $0.95$, leading to
$\G(8.5, 7.5)$. To obtain a lower bound of $\tau_\epsilon$ we assumed a 
uniform distribution of $\bm{y}$ in $[0,1]$, because all populations are 
growing in the absence of density-dependent effects ($\bm{y}>0$) and their 
growth is restricted by the number of offspring per animal and year ($\bm{y}<1$).
As upper bound we used 100 divided by the
sample variance of $\bm{y}$, so that the coefficient of determination
is $R^2=0.99$. This leads to $\G(1, 0.001)$.

\begin{figure}
\centering
\mbox{\includegraphics[width=14cm]{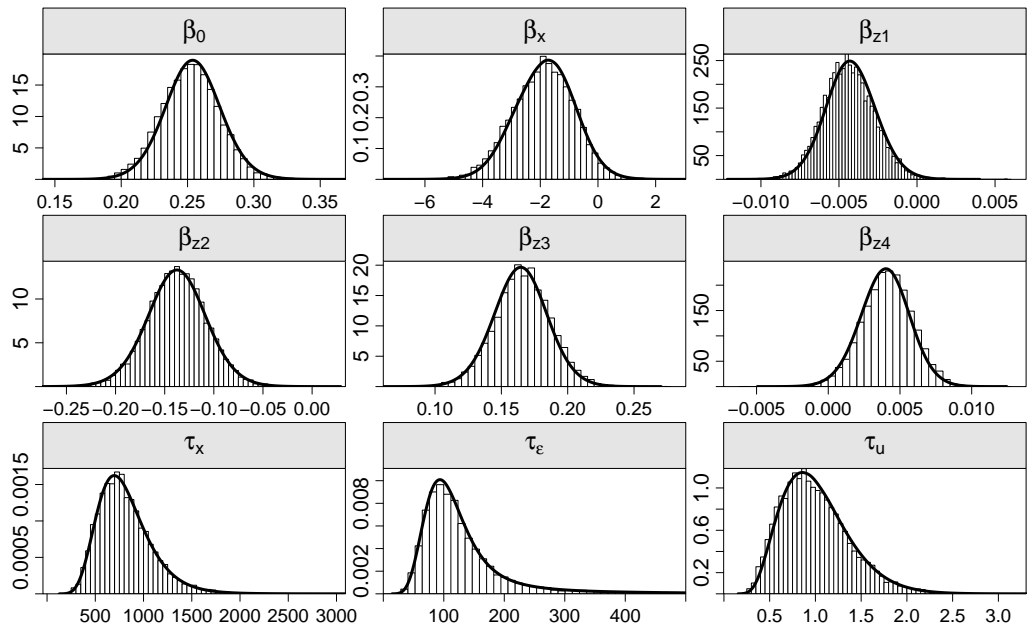}}
\caption{\label{IbexHist}Comparison of the MCMC samples (histograms) with the 
INLA posterior marginals (lines) for the ibex data. The $\bm{z}$-covariates are those 
treated as error-free. They are the length of the time series ($\bm{z}_1$), 
average precipitation in summer ($\bm{z}_2$), average precipitation in 
winter ($\bm{z}_3$) and the interaction $\bm{z}_4=\bm{z}_1\bm{z}_2$.}
\end{figure}

\begin{figure}
\centering
\mbox{\includegraphics[width=\textwidth]{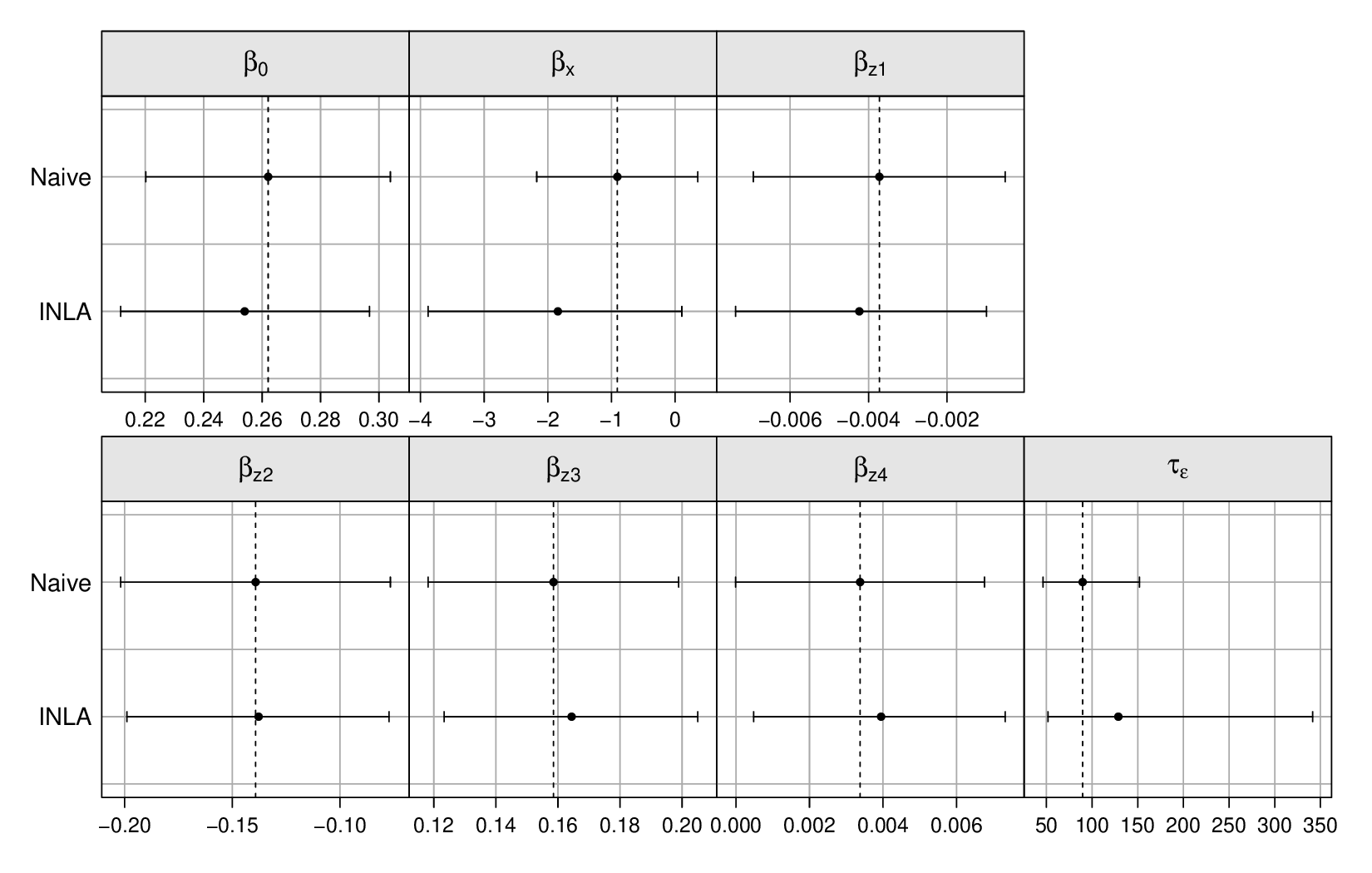}}
\caption{\label{IbexTable}Posterior means and 95\% credible intervals in the ibex data analysis. 
The dashed lines indicate the naive estimates. Only parameters where a naive 
estimate was available are included here. }
\end{figure}

An MCMC simulation was run for \numprint{100000} iterations with a burn-in of 
\numprint{10000} iterations and a saving frequency of 10. The estimates obtained 
from INLA were chosen as starting values. Convergence was visually checked. 
Figure~\ref{IbexHist} shows the perfect fit between the MCMC samples and the 
posterior marginals of INLA. Of note, due to the Gaussian likelihood the 
results obtained by INLA are exact and contain no approximation error.
The parameter estimates are graphically compared to the naive Bayesian analysis in Figure~\ref{IbexTable}, 
including $\bm{w}$ instead of $\bm{x}$ and using the same priors for the respective parameters. 
The absolute value of the slope $|\beta_x|$ and the residual precision 
$\tau_\eps$ are underestimated in this naive regression, as predicted by 
the theory. The other parameters are only slightly affected by the error in $\bm{x}$.

\subsection{Influence of systolic blood pressure on coronary heart disease}\label{Framingham}
The outcome $y_i \in [0,1]$ in this study is an indicator for coronary heart
disease, assumed to be Bernoulli distributed. 
The observation model is logistic, using an indicator for smoking, $\bm{z}$, and 
the transformed (unobserved) long-term blood pressure $\bm{x} = \log(\text{SBP} - 50)$ as 
binary and continuous covariates, respectively.
Hence, the linear predictor is 
\begin{equation*}
\logit\left[\P(\bm{y}=1 \, | \, \bm{x}, \bm{z})\right] =\beta_0\mm{1}+\beta_x \bm{x} + \beta_z\bm{z} \ .\label{FramRegrModel}
\end{equation*}
Since the SBP has been measured at two different examinations, the magnitude of the measurement
error of these surrogate measures can be quantified.  Here, we assume that the
repeated measurements $\bm{w}_1$ and $\bm{w}_2$ at examination 1 and 2,
respectively, are independent and normally distributed with mean $\bm{x}$ and
precision $\tau_u$, leading to the classical homoscedastic error model
\begin{equation*}
  \bm{w} \given \bm{x} \sim \mathcal{N}\left(\begin{pmatrix}\bm{x}\\\bm{x}\end{pmatrix}, \tau_u \mathbf{I}\right) \ ,
\end{equation*}
where $\bm{w} = (\bm{w}_1^\top, \bm{w}_2^\top)^\top$, and $\mathbf{I}$ is of dimension $2n \times 2n$ with $n=641$.

Finally, the exposure model~\eqref{eq:ExpModel} comes in its most general form
\begin{equation*}
\bm{x} \given \bm{z} \sim \mathcal{N}(\alpha_0\mm{1}+ \alpha_z\bm{z}, \tau_{x}\mathbf{I}) \ . 
\end{equation*}

The latent field in this model is $\bm{v}=(\bm{x}^\top,\beta_0, \beta_z, \alpha_0, \alpha_z)^\top$,
and the hyperparameters are $\bm{\theta}=(\beta_x,\tau_u,\tau_{x})^\top$.

For $\beta_0$, $\beta_x$ and $\beta_z$ we assigned independent 
$\mathcal{N}(0,10^{-2})$ priors. The remaining prior distributions are specified
based on prior considerations. We assume that $90$ mmHg and $180$ mmHg can be regarded
as the respective $2.5\%$ and $97.5\%$ quantile of SBP, and that 
$\text{SBP} - 50 \sim \text{LogNormal}(\mu, \sigma^2)$. Through optimization we
determined $\mu \approx 4.3$ and 
$\sigma^2 \approx 0.1$, so that 
the log normal distribution has the desired quantiles. Consequently, we used
$1/\sigma^2$ as expected value
for $\tau_x$. Assuming equal mean and variance for $\tau_x$ we specified 
$\tau_x \sim \G(10,1)$, and further $\alpha_0 \sim \mathcal{N}(0, 1)$, whereas $\mu=0$ is used instead of $\mu=4.3$ due to the centering of $\bm{w}_1$ and $\bm{w}_2$.
\cite{rothe.kim1980} found the measurement error of SBP 
to be as much as $20$ mmHg, meaning that our assumed mean SBP of $135$ mmHg varies between
$115$ and $155$. This corresponds to an error factor of $1.15$, from which 
we derive an expected value of approximately $100$ for $\tau_u$. 
Assuming again equal mean and variance of the prior for the precision, we set
$\tau_u \sim \G(100,1)$. For $\alpha_z$ we assume a mean of zero, and set
$\alpha_z \sim \mathcal{N}(0,1)$.
%
Note, that these prior specifications might deviate from the reference example in \cite{carroll.etal2006}, 
where the exact parameters were not given.
Furthermore, \cite{carroll.etal2006} used the quantity $\Delta:=\tau_{x}/\tau_u$ 
instead of $\tau_u$, and gave it a uniform prior in the interval $(0,0.5)$. 
Since this is not straightforward with INLA, the model was modified as described.

To obtain Markov Chain Monte Carlo (MCMC) posterior marginals, regression 
coefficients of GLMs cannot directly be sampled from a standard full 
conditional distribution. Here, samples were obtained according 
to \cite{Gamerman1997}. The algorithm can be used if the 
observations $y_i$ are conditionally independent and follow an 
exponential family density. For the regression coefficients 
$\bm\beta=(\beta_0,\beta_x,\bm\beta^\top_z)^\top$, this approach uses 
transition densities that combine the weighted least squares method with a 
prior on $\bm{\beta}$ \citep{McCullaghNelder1989, West1985}.
The full conditionals for the unknowns in our logistic regression model 
are given in Section~8.1 of the Supplementary Material.

The simulation was run for \numprint{100000} iterations with a 
burn-in of \numprint{10000}, and every 5$^{th}$ value was saved. 
Starting values for $\bm\alpha$ and $\bm\beta$ were
chosen from the INLA output. For $\tau_u$ and $\tau_{x}$, the mean
of their respective prior distribution were used as initial estimates.

\begin{figure}
\centering
\mbox{\includegraphics[width=\textwidth]{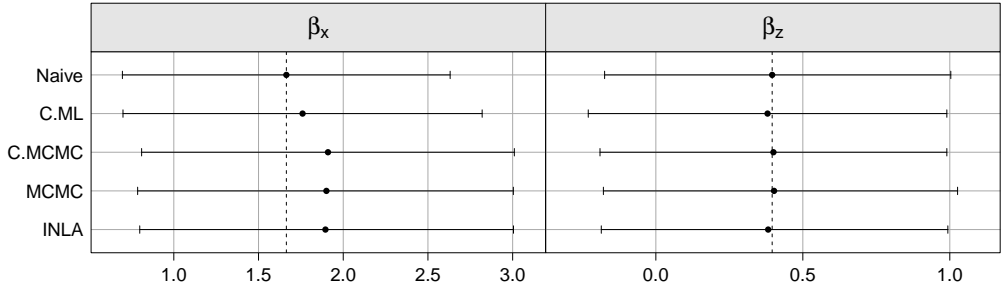}}
\caption{\label{FramTable}Posterior means and 95\% credible intervals for the Framingham data 
analysis. For MCMC and INLA, posterior means are used as point estimates. 
\emph{C.MCMC} and \emph{C.ML} stand for the Bayesian and the maximum 
likelihood analysis conducted in \citet{carroll.etal2006}. The dashed 
lines indicate the naive estimates.}
\end{figure}

The agreement between the MCMC and INLA output is almost perfect, compare supplementary Figure~1. 
Figure~\ref{FramTable} shows parameter estimates for $\beta_x$ and $\beta_z$
obtained by the naive regression model including $\bm{w}_1$ and $\bm{w}_2$
instead of $\bm{x}$, and four error-correction approaches. 
\cite{carroll.etal2006} used a measurement error model fitted via 
a maximum-likelihood method and a Bayesian approach using MCMC, 
denoted here as C.ML and C.MCMC. 
The fourth and fifth rows show the results obtained by our MCMC implementation 
and INLA. All error-corrected estimates and the credible intervals are similar.
While the coefficient $\beta_z$ of the error-free measured smoking
status seems unbiased, the effect of systolic blood pressure is clearly 
attenuated in the naive analysis. Adjusting for measurement error leads to 
a more pronounced effect, as expected however with a larger assigned uncertainty.

\subsection{Seedling growth across different light conditions}\label{Berkson}

Let $\bm{y}$ denote the number of new leaves per plant after a four months 
growth phase. The covariate $\bm{z}$ denotes the degree of defoliation and 
$\bm{x}=\log(\% \text{light})$  the (transformed) light intensity, where $\bm{w}$ is the 
respective target value. 
Using $\bm{w}$ instead of $\bm{x}$ in the analysis
leads to the homoscedastic Berkson error with
\begin{equation*}
 \bm{x} \, | \,  \bm{w} \sim \mathcal{N}( \bm{w},\tau_u\mathbf{I}) \ .
\end{equation*}
In the following we centered both covariates $\bm{w}$ and $\bm{z}$.  
This data structure leads to a Poisson regression model with nested design. 
To account for overdispersion, independent normal random effects 
$\gamma_{ijk}\sim \mathcal{N}(0,\tau_\gamma)$ were added, extending the GLM to a GLMM:
\begin{equation*}\label{eq:shading}
\log(\E(y_{ijk} \given \bm{x}, \bm{z}, \bm{\beta}, \bm{\gamma}))=\beta_0 + \beta_x x_{ij} + \beta_z z_k + \gamma_{ijk}\ ,
\end{equation*}
with $i=1, 2, 3$ denoting the light condition, $j=1,\ldots, 5$ the shadehouse per light condition, and $k=1,\ldots, 4$ the degrees of defoliation. The unknowns of this model are $\bm{v}=(\bm{x}^\top,\beta_0,\beta_z)^\top$ and $\bm{\theta}=(\beta_x,\tau_u,\tau_\gamma)^\top$.

The $\bm{\beta}$ parameters were assigned independent $\mathcal{N}(0, 10^{-2})$
priors, and the overdispersion precision $\tau_\gamma$ a highly dispersed but
proper $\text{G}(1,0.005)$ prior with mean 200.  For the error precision
$\tau_u$ it was assumed that the actual light values $\bm{x}$ do not interfere
with the target values $\bm{w}$ from other light levels. The (centered and
log-transformed) target light values are 1.22, 0.10 and -1.32 for dark, middle
and light conditions, thus the interval between middle and light measurements is
1.42. Interpreting this as one branch of a 95\% confidence interval of a
Gaussian variable, we obtain $\sigma_u=1.42/1.96=0.72$, yielding a lower bound
for $\tau_u$ of $1/0.72^2=1.93$. For the upper bound ten times less variation is
assumed, leading to an upper limit of $1/0.072^2=193$.  The gamma distribution
with the respective 2.5\% and 97.5\% quantiles is $\tau_u\sim\G(1,0.02)$.

\begin{figure}
\centering
\mbox{\includegraphics[width=14cm]{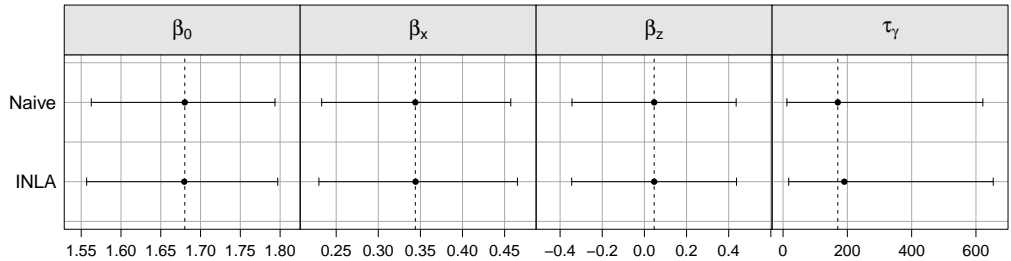}}
\caption{\label{ShadingHist}Posterior means and 95\% credible intervals for the seedling 
growth example. The dashed lines indicate the naive estimates. 
Only parameters where a naive estimate was available are included here.}
\end{figure}

 
The results from the regression with INLA were compared to an MCMC run with \numprint{100000} 
iterations, a burn-in of \numprint{10000} iterations, and a saving frequency of 10. Sampling was based 
on a reparameterization as proposed by~\cite{BesagGreenHigdonMengersen1995}, 
where all except one full conditional distribution are standard and can be Gibbs-sampled. 
The MCMC samples and posterior marginals fit very well, see supplementary Figure~2.
The parameter estimates from the naive analysis including $\bm{w}$ and the error-corrected estimates 
of INLA are shown in Figure~\ref{ShadingHist}. As mentioned in the introductory 
Section~\ref{ShadingIntro}, we observe no bias in the regression coefficients, yet there is a 
small bias in the precision of the random effect $\tau_\gamma$. Moreover, the credible 
intervals for $\beta_0$, $\beta_x$ and $\tau_\gamma$ are slightly increased. Note that
the same framework as presented here can be used for logistic regression models, where
Berkson error is known to cause bias in the parameter estimates 
\citep{Burr1988, bateson.etal2010}. 



\section{Discussion}\label{sec:discussion}
Measurement error in covariates may lead to serious biases in parameter 
estimates and confidence intervals of statistical models. 
A variety of approaches to model such error have been proposed in the past decades, 
among which Bayesian methods probably provide the most flexible framework. 
Bayesian treatments, employing MCMC samplers, have been successfully applied 
for more than 20 years, but their application has never become part of standard regression analyses. 

The aim of this work was to illustrate how the most common \ac{ME} models
(classical and Berkson error) can be included in GLMMs using the recently
proposed INLA framework, which gives fast and accurate approximations instead of
doing any sampling.  The provided R-code should help to make such models
accessible to a broader audience. Note that INLA provides a much larger variety
of likelihood functions and latent models than we could illustrate here, and the
modular structure adds to its flexibility.  It is, for instance, straightforward
to treat several mismeasured covariates jointly, to introduce a systematic bias
into the error model, or to include any structured random term into the model
formulation.  Gaussian classical and Berkson error models naturally fit into the
INLA framework of latent Gaussian models, and thus the error-prone covariates
used here are always continuous.  

The treatment of more general error models is also possible.  One interesting
application, relevant for example in ecology, is the use of non-Gaussian error
models, for example a Poisson or negative binomial model, where instead of the
true and positive (but unobserved) continuous covariate $\bm{x}$, a discrete
proxy $\bm{w}$ with mean $\bm{x}$ is observed. More general models might also be
useful, e.g.~a log-linear model with mean $\E(\bm{w})=a \bm{x}^{\beta_x}$ or a
logistic model for binomial proxies.  Furthermore, it might not always be
appropriate to assume that the components of $\bm{x}$ are iid.  Hence, $\bm{x}$
could follow a more complex Gaussian Markov random field structure
\citep{GMRFbook} to account for temporal and/or spatial dependencies, see
\cite{bernardinelli.etal1997} for such a formulation in an epidemiological
context. Both of these extensions can be handled with INLA and will be detailed
in future work.

One of the biggest challenges when treating mismeasured variables is the 
estimation of the error variance, either from repeated measurements, 
instrumental variables or from previous studies. The advantage of a Bayesian 
approach, as the one taken here, is that uncertainty of such estimates can 
be incorporated into prior distributions. 
Sensitivity to chosen prior assumptions can be easily checked due to the 
computational speed of INLA, see \citet{roos.held2011}.


\section*{Supplementary Material for ``Measurement error in GLMMs with INLA''}
Due to space constraints, the R-code for all examples presented here is 
described in detail in the supplementary document. Furthermore, this 
document contains full conditionals and posterior marginals for 
Section~\ref{Framingham}. On \url{www.r-inla.org/examples/case-studies/muff-etal-2013} 
selected data and R-code are provided for download.

\section*{Acknowledgements}
We thank Lukas Keller for stimulating the ME project and many 
interesting discussions. We are grateful to Lukas Keller, Iris Biebach and 
Claudio Bozzuto for providing us with the ibex data. 
We also thank Helmut K\"uchenhoff and Robert Bagchi for useful 
discussions and suggestions to this work, and Malgorzata Roos for comments to the manuscript.

\bibliographystyle{Chicago}
\bibliography{literatur}

\clearpage

\section*{\LARGE Supplementary Material for ``Bayesian analysis of measurement error models using INLA"\\[0.5cm]}

\setcounter{section}{6}
\setcounter{equation}{9}

\section{R-code for the three applications in the main text}
In this section we guide the reader through the {\tt r-inla} code and technical details of the three examples discussed in the main text.  On \url{www.r-inla.org/examples/case-studies/muff-etal-2013} selected data and R-code are provided for download. The {\tt r-inla} package can be installed by typing the following command line in the {\tt R} terminal:

\begin{Schunk}
\begin{Sinput}
 source("http://www.math.ntnu.no/inla/givemeINLA.R")
 upgrade.inla(testing=TRUE)
\end{Sinput}
\end{Schunk}
Using
\begin{Schunk}
\begin{Sinput}
 inla.version()
\end{Sinput}
\end{Schunk}
information regarding the actual installed version is shown. Here, we used the {\tt r-inla} version built on July 13, 2013. For more information regarding the installation process we refer to \url{www.r-inla.org}.

\subsection{Inbreeding in Swiss ibex populations (classical error)}\label{HeteroApp}
Let all variables be defined as in Section 5.1 of the main text. Recall that the model is Gaussian and contains five covariates ($\bm{x}, \bm{z}_1, \bm{z}_2, \bm{z}_3$ and $\bm{z}_4$). The covariate $\bm{x}$ is not directly observed, but only a proxy $\bm{w}$ following a classical heteroscedastic error model $\bm{w}\given\bm{x}\sim\mathcal{N}(\bm{x},\tau_u\mathbf{D})$.
The prior distributions are elicited from expert/prior knowledge, see main text, 
and are defined as:
\begin{itemize}
\item $\bm{x} \sim \mathcal{N}(\mm{0},\tau_x\mathbf{I})$.
\item $\beta_0, \beta_x, \beta_{z_1}, \beta_{z_2}, \beta_{z_3}, \beta_{z_4} \sim \mathcal{N}(0, \tau_\beta)$, with $\tau_\beta=0.0001$, 
\item $\tau_x \sim \G(\alpha_x, \beta_x)$, with $\alpha_x= 1$ and $\beta_x=0.0009$,
\item $\tau_y \sim \G(\alpha_y, \beta_y)$, with $\alpha_x= 1$ and $\beta_x=0.001$,
\item $\tau_u \sim \G(\alpha_u, \beta_u)$, with $\alpha_u = 8.5$ and $\beta_u = 7.5$.
\end{itemize}

\nin
The object {\tt data} consists of seven columns:
$$\bm{y} \, \quad \bm{w} \,\quad  \bm{z}_1 \, \quad  \bm{z}_2 \, \quad  \bm{z}_3 \, \quad  \bm{z}_4 \, \quad \text{\bf error.prec}$$
They contain (for $n=26$):
\begin{itemize}
\item $y_1\ldots y_{n}$: The populations' intrinsic growth rates.
\item $w_1 \ldots w_{n}$: The estimated inbreeding coefficients (proxies for $x_1 \ldots x_n$; centered).
\item $z_{11} \ldots, z_{1n}$: Length of the time series (centered).
\item $z_{21} \ldots, z_{2n}$: Average precipitation in summer (centered).
\item $z_{31} \ldots, z_{3n}$: Average precipitation in winter (centered).
\item $z_{41} \ldots, z_{4n}$: Interaction between $\bm{z}_1$ and $\bm{z}_2$.
\item $\text{error.prec}_1 \ldots \text{error.prec}_{n}$: The error precisions in the estimates $\bm{w}$.
\end{itemize}

Start with the prior specification process as described above and in the main text:

\begin{Schunk}
\begin{Sinput}
 data <- read.table("ibex_data4supp.txt", header=T)
 attach(data)
\end{Sinput}
\end{Schunk}
\begin{Schunk}
\begin{Sinput}
 prior.beta <- c(0, 0.0001)
 prior.prec.x <- c(1, 0.0009)
 prior.prec.y <- c(1, 0.001)
 prior.prec.u <- c(8.5, 7.5)
\end{Sinput}
\end{Schunk}
\begin{Schunk}
\begin{Sinput}
 # initial values (mean or mode of prior) 
 prec.x = 1/0.0009  
 prec.y = 1/0.001 
 prec.u = 1
\end{Sinput}
\end{Schunk}
\nin
Next, we define the INLA model formula. There are four fixed effects ($\beta_{z_1}$, $\beta_{z_2}$, $\beta_{z_3}$, $\beta_{z_4}$) and one random effect $\beta_x$ belonging to the error-prone covariate $\bm{x}$, where the new {\tt mec} model is employed for the latter. Note that the heteroscedasticity in the error in $\bm{w}$ is encoded by assigning  the vector of error precisions {\tt error.prec} to the {\tt scale} option. In the {\tt values} option, all values of $\bm{w}$ must be listed. The model contains four hyperparameters:
\begin{itemize}
\item {\tt beta} corresponds to $\beta_x$, the slope coefficient of the error-prone covariate $\bm{x}$, with a Gaussian prior.
\item {\tt prec.u} is the error precision $\tau_u$ with gamma prior.
\item {\tt prec.x} is the precision $\tau_x$ of $\bm{x}\sim  \mathcal{N}(\alpha_0 \bm{1}, \tau_x \mathbf{I})$ with gamma prior.
\item {\tt mean.x} corresponds to the mean $\alpha_0$, which is fixed here at $0$ due to covariate centering.
\end{itemize}
The prior settings are defined in the different entries of the list {\tt hyper}. The option {\tt fixed} specifies whether the corresponding quantity should be estimated or fixed at the {\tt initial} value. The field {\tt param} captures the prior parameters of the corresponding prior distribution. Gaussian prior distributions are the default for ${\tt beta}$ and ${\tt mean.x}$, while log-gamma distributions are used for the log-transformed precisions ${\tt prec.u}$ and ${\tt prec.x}$. Note hereby that if a variable $\tau$ is gamma distributed with shape parameter $a$ and rate parameter $b$ leading to the mean $a/b$ and variance $a/b^2$, then $\log(\tau)$ is log-gamma distributed with the same parameters $a$ and $b$.

\begin{Schunk}
\begin{Sinput}
 library(INLA)
 formula <- y ~ f(w, model = "mec", scale = error.prec, values = w, 
   hyper = list(
    beta = list(
      param = prior.beta,
      fixed = FALSE
      ),
    prec.u = list(
      param = prior.prec.u,
      initial = log(prec.u),
      fixed = FALSE
      ),
    prec.x = list(
      param = prior.prec.x,
      initial = log(prec.x),
      fixed = FALSE
      ),
    mean.x = list(
      initial = 0,
      fixed = TRUE
      )
    )
  ) + z1 + z2 + z3 + z4
\end{Sinput}
\end{Schunk}
\nin
The call of the {\tt inla} function includes the specifications for $\tau_\varepsilon$, the hyperparameter of the Gaussian regression model. These can be controlled via the {\tt control.family} option. The prior distributions for the intercept $\beta_0$ and the fixed effects of the other covariates $\bm{z}_1, \ldots, \bm{z}_4$ are specified in the {\tt control.fixed} option.

\begin{Schunk}
\begin{Sinput}
 r <- inla(formula, data = data.frame(y, w, z1, z2, z3, z4, error.prec),
           family = "gaussian",
           control.family = list(
               hyper = list(
                  prec = list(param = prior.prec.y,
                           initial = log(prec.y),
                           fixed = FALSE
                  )
               )
            ),
           control.fixed = list(
             mean.intercept = prior.beta[1],
             prec.intercept = prior.beta[2],
             mean = prior.beta[1],
             prec = prior.beta[2]
           )
      )
 r <- inla.hyperpar(r, dz = 0.5, diff.logdens = 20)
\end{Sinput}
\end{Schunk}
The last command improves the estimates of the posterior marginals for the hyperparameters of the model. The call is optional, but a slightly better agreement with the MCMC posterior marginals was found in this example. To get a quick overview of the results, use the {\tt summary} command.
\begin{Schunk}
\begin{Sinput}
 summary(r)
\end{Sinput}
\end{Schunk}
\vspace{1cm}

\subsection{Influence of systolic blood pressure on coronary heart disease (classical error)}\label{FramApp}
\nin
Let all variables be defined as in Section 5.2 of the main text. 
The outcome is binary in $[0,1]$, and assumed to be binomial distributed, 
i.e.~$\p(y_i)={N \choose y_i} \pi_i^{y_i} (1-\pi_i)^{1-y_i}$, with
$N=1$, $\pi_i= \exp(\eta_i)/(1+\exp(\eta_i))$  and
and $\eta_i=\beta_0+\beta_xx_i+\beta_zz_i$. We have a classical error structure, 
where the covariate $\bm{x}=(x_1, \ldots, x_n)^\top$ is not directly observed, 
but two replicates, $\bm{w}_1=(w_{11}, \ldots, w_{1n})^\top$ and 
$\bm{w}_2=(w_{21}, \ldots, w_{2n})^\top$ are used as proxy, where 
$\bm{w}_1 \sim \mathcal{N}(\bm{x}, \tau_u\mathbf{I})$ and $\bm{w}_2 \sim \mathcal{N}(\bm{x}, \tau_u\mathbf{I})$. 
The prior distributions are elicited from expert/prior knowledge, see main text, 
and are defined as:
\begin{itemize}
\item $\bm{x} \sim \mathcal{N}(\alpha_0 + \alpha_z \bm{z}, \tau_x\mathbf{I})$.
\item $\beta_0, \beta_x, \beta_z \sim \mathcal{N}(0, \tau_\beta)$, with $\tau_\beta=0.01$, 
\item $\alpha_0 \sim \mathcal{N}(\mu_{\alpha_0}, \tau_{\alpha_0})$, with $\mu_{\alpha_0} = 0$ and $\tau_{\alpha_0} = 1$, 
\item $\alpha_z \sim \mathcal{N}(\mu_{\alpha_z}, \tau_{\alpha_z})$, with $\mu_{\alpha_z}=0$ and $\tau_{\alpha_z} = 1$.
\item $\tau_x \sim \G(\alpha_x, \beta_x)$, with $\alpha_x= 10$ and $\beta_x=1$,
\item $\tau_u \sim \G(\alpha_u, \beta_u)$, with $\alpha_u = 100$ and $\beta_u = 1$.
\end{itemize}

\nin
The object {\tt data} consists of four columns:
$$\bm{y} \, \quad \bm{w_1} \,\quad \bm{w_2} \,\quad  \bm{z}$$
They contain (for $n=641$):
\begin{itemize}
\item $y_1\ldots y_{n}$: The binary response $y_i\in\{0,1\}$.
\item $w_{11} \ldots w_{1n}$: $\log(\text{SBP}-50)$ at examination 1 (centered).
\item $w_{21} \ldots w_{2n}$: $\log(\text{SBP}-50)$ at examination 2 (centered).
\item $z_{1} \ldots z_{n}$: Smoking status $z_i\in\{0,1\}$.
\end{itemize}
As described in the main text, the hierarchical model of this example is formulated in INLA as a joint model by applying the {\tt copy} feature. The full model can be written as

\begin{equation*}\label{jointModel}
\underbrace{\begin{bmatrix}
y_{1} & \na & \na \\
\vdots & \vdots & \vdots \\
y_{n} & \na & \na \\
\na & 0 & \na\\
\vdots & \vdots & \vdots \\
\na & 0 & \na\\
\na & \na & w_{11} \\
\vdots & \vdots & \vdots \\
\na & \na & w_{1n} \\
\na & \na & w_{21} \\
\vdots & \vdots & \vdots \\
\na & \na & w_{2n} \\
\end{bmatrix}}_{ {\tt Y}}
=  \beta_0   \underbrace{\begin{bmatrix} 1 \\ \vdots \\ 1\\ \na \\ \vdots \\ \na \\ \na \\ \vdots   \\ \na \\ \na \\ \vdots   \\ \na\end{bmatrix}}_{ {\tt beta.0}}
+ \beta_x   \underbrace{\begin{bmatrix} 1 \\ \vdots \\ n\\ \na \\ \vdots \\ \na \\ \na \\ \vdots   \\ \na\\ \na \\ \vdots   \\ \na \end{bmatrix} }_{ {\tt beta.x}}
+ \underbrace{\begin{bmatrix} \na \\ \vdots \\ \na \\ -1 \\ \vdots \\ -n \\1 \\  \vdots \\ n\\1 \\ \vdots \\ n \end{bmatrix}}_{{\tt idx.x}}
+ \beta_z   \underbrace{\begin{bmatrix} z_1 \\ \vdots \\ z_n\\ \na \\ \vdots \\ \na \\ \na \\ \vdots   \\ \na\\ \na \\ \vdots   \\ \na \end{bmatrix}}_{{\tt beta.z}}  
+ \alpha_0 \underbrace{\begin{bmatrix} \na \\ \vdots \\ \na \\ 1 \\ \vdots \\ 1 \\ \na \\  \vdots \\ \na\\ \na \\ \vdots   \\ \na \end{bmatrix}}_{ {\tt alpha.0}}
+ \alpha_z \underbrace{\begin{bmatrix} \na \\ \vdots \\ \na \\ z_1 \\ \vdots \\ z_n \\ \na \\  \vdots \\ \na\\ \na \\ \vdots   \\ \na \end{bmatrix}}_{ {\tt alpha.z}} \ .
\end{equation*}

The reader is guided through the {\tt r-inla} code for this joint model formulation in the following. The terms below the brackets indicate the names as they will be employed in the code. 
Start with the prior specification process, as described in the main text:
\begin{Schunk}
\begin{Sinput}
 data <- read.table("fram_data4supp.txt", header=T)
 attach(data)
 n <- nrow(data)  #641
\end{Sinput}
\end{Schunk}
\begin{Schunk}
\begin{Sinput}
 prior.beta <- c(0, 0.01)
 prior.alpha0 <- c(0, 1)
 prior.alphaz <- c(0, 1)
 prior.prec.x <- c(10, 1)
 prior.prec.u <- c(100, 1)
\end{Sinput}
\end{Schunk}
\begin{Schunk}
\begin{Sinput}
 # initial values (mean of prior)
 prec.u <- 100
 prec.x <- 10
\end{Sinput}
\end{Schunk}
Second, the response matrix {\tt Y} and the data vectors are filled according 
to the naming of the above joint model equation:

\begin{Schunk}
\begin{Sinput}
 Y <- matrix(NA, 4*n, 3)
 Y[1:n, 1] <- y
 Y[n+(1:n), 2] <- rep(0, n)
 Y[2*n+(1:n), 3] <- w1
 Y[3*n+(1:n), 3] <- w2
 beta.0 <- c(rep(1, n), rep(NA, n), rep(NA, n), rep(NA, n))
 beta.x <- c(1:n, rep(NA, n), rep(NA, n), rep(NA, n))
 idx.x <- c(rep(NA, n), 1:n, 1:n, 1:n)
 weight.x <- c(rep(1, n), rep(-1, n), rep(1, n), rep(1,n))
 beta.z <- c(z, rep(NA, n), rep(NA, n), rep(NA,n))
 alpha.0 <- c(rep(NA, n), rep(1, n), rep(NA, n), rep(NA, n))
 alpha.z <- c(rep(NA, n), z, rep(NA, n), rep(NA, n))
 Ntrials <- c(rep(1, n), rep(NA, n), rep(NA, n), rep(NA, n))
 data.joint <- data.frame(Y=Y, 
      beta.0=beta.0, beta.x=beta.x, beta.z=beta.z,
      idx.x=idx.x, weight.x=weight.x,
      alpha0=alpha.0, alpha.z=alpha.z,
      Ntrials=Ntrials)
\end{Sinput}
\end{Schunk}

The next step contains the definition of the INLA formula. 
There are four fixed effects ($\beta_0$, $\beta_z$, $\alpha_0$ and $\alpha_z$) 
and two random effects. The latter are needed to encode that the 
values of $\bm{x}$ in the exposure~(7) and error model~(8) are assigned the 
same values as in the regression model~(6), where $\beta_x\bm{x}$ represents 
a product of two unknown quantities. The two random effects terms are:
\begin{itemize}
\item {\tt f(beta.x,...)}: The {\tt copy="idx.x"} call guarantees the 
  assignment of identical values to $\bm{x}$ in all components of the 
  joint model. As discussed in the main text, $\beta_x$ is treated as a 
  hyperparameter, namely the scaling parameter of the copied process 
   $\bm{x}^\star$. 
\item {\tt f(idx.x,...)} : {\tt idx.x} contains the $\bm{x}$ values, 
  encoded as an i.i.d.\ Gaussian random effect, and weighted with 
  {\tt weight.x} to ensure correct signs in the joint model. 
  The {\tt values} option contains the vector of all values assumed by 
  the covariate for which the effect is estimated. It must be a numeric 
  vector, a vector of factors or NULL.
  The precision {\tt prec} of the random effect is fixed at $\tau=\exp(-15)$. 
  This is necessary as the uncertainty in $\bm{x}$ is already modelled in
  the second level (column 2 of $\mathbf{Y}$) of the joint model, which 
  defines the exposure component.
\end{itemize}

\begin{Schunk}
\begin{Sinput}
 library(INLA)
 formula <- Y ~  f(beta.x, copy = "idx.x",
      hyper = list(beta = list(param = prior.beta, fixed = FALSE))) +
    f(idx.x, weight.x, model = "iid", values = 1:n,
      hyper = list(prec = list(initial = -15, fixed = TRUE))) +
    beta.0 - 1 + beta.z + alpha.0 + alpha.z
\end{Sinput}
\end{Schunk}
Since there is no common intercept in the joint model, it has to be explicitly 
removed using {\tt -1}.
The call of the {\tt inla} function is given next. The following options 
need some explanation:
\begin{itemize}
\item {\tt family}: There are three different likelihoods here, namely the 
binomial likelihood of the regression model and two Gaussian likelihoods, 
one for the exposure and one for the error model. They correspond to the 
different columns in the response matrix {\tt Y}. 
\item {\tt control.family}: Specification of the hyperparameters for the 
three likelihoods, in the same order as given in {\tt family}. The binomial 
likelihood does not contain any hyperparameters, thus the respective list is 
empty. In the second and third likelihoods the hyperparameters $\tau_x$ and 
$\tau_u$ need to be specified, respectively.
\item {\tt control.fixed}: Prior specification for the fixed effects.
\end{itemize}
\begin{Schunk}
\begin{Sinput}
 r <- inla(formula, Ntrials = Ntrials, data = data.joint,
    family = c("binomial", "gaussian", "gaussian"),
    control.family = list(
        list(hyper = list()),
        list(hyper = list(
  	  prec = list(initial = log(prec.x),
  	  param = prior.prec.x,
  	  fixed = FALSE))),
        list(hyper = list(
  	  prec = list(initial=log(prec.u),
  	  param = prior.prec.u,
  	  fixed = FALSE)))),
    control.fixed = list(
        mean = list(beta.0=prior.beta[1], beta.z=prior.beta[1], 
  	  alpha.z=prior.alphaz[1], alpha.0=prior.alpha0[1]),
        prec = list(beta.0=prior.beta[2], beta.z=prior.beta[2], 
  	  alpha.z=prior.alphaz[2], alpha.0=prior.alpha0[2]))
  )
 r <-inla.hyperpar(r)
\end{Sinput}
\end{Schunk}
The last call ({\tt inla.hyperpar}) is not required. It is used to improve 
the estimates of the posterior marginals for the hyperparameters 
using a finer grid in the numerical integration. In this application, only
the marginal of $\tau_x$ changes slightly by this correction.\\

\subsection{Seedling growth across different light conditions (Berkson error)}\label{HeteroApp}

Let all variables be defined as in Section 5.3 of the main text. Recall that the model is Poisson and contains the two covariates $\bm{x}$ and $\bm{z}$, and one independent, normal random effect $\bm{\gamma}$ to account for potential overdispersion. The covariate $\bm{x}$ is not directly observed, but only a proxy $\bm{w}$ following a Berkson error model $\bm{x}\given \bm{w}\sim\mathcal{N}(\bm{w},\tau_u\mathbf{I})$.
The prior distributions are elicited from expert/prior knowledge, see main text, 
and are defined as:
\begin{itemize}
\item $\beta_0, \beta_x, \beta_{z} \sim \mathcal{N}(0, \tau_\beta)$, with $\tau_\beta=0.01$, 
\item $\tau_\gamma \sim \G(\alpha_\gamma, \beta_\gamma)$, with $\alpha_\gamma = 1$ and $\beta_\gamma = 0.005$,
\item $\tau_u \sim \G(\alpha_u, \beta_u)$, with $\alpha_u = 1$ and $\beta_u = 0.02$.

\end{itemize}
\vspace{5mm}

\noindent{\bf  Analysis with the {\tt meb} model}\\[-4mm]

\nin
The object {\tt data} consists of three columns:
$$\bm{y} \, \quad \bm{w} \,\quad \bm{z}$$
They contain (for $n=60$):
\begin{itemize}
\item $y_1\ldots y_{n}$: The number of new leaves.
\item $w_{1} \ldots w_{n}$: $\log(\%\text{light})$ for the target light intensities under dark, middle and light conditions (i.e., only three different values; centered).
\item $z_{1} \ldots z_{n}$: Degree of defoliation (0\%, 25\%, 50\%, 75\%; centered).
\end{itemize}

Let us start again with prior specification process in accordance to the main text:
\begin{Schunk}
\begin{Sinput}
 data <- read.table("shading_data4supp.txt", header=T)
 attach(data)
 n <- 60 	  # number of seedlings
 s <- 15 	  # number of shadehouses
 w <- w + rep(rnorm(s,0,1e-4),each=n/s)
 individual <- 1:n # id to incorporate individual random effects
\end{Sinput}
\end{Schunk}

\begin{Schunk}
\begin{Sinput}
 prior.beta <- c(0,0.01)
 prior.tau <- c(1,0.005)   
 prior.prec.u <- c(1,0.02)   
\end{Sinput}
\end{Schunk}
\begin{Schunk}
\begin{Sinput}
 # initial values (mean of prior)
 prec.tau <- 1/0.005 
 prec.u <- 1/0.02    
\end{Sinput}
\end{Schunk}

The fourth line contains a trick to ensure that the light values ${\tt w}$ from the $s=15$ shadehouses are not completely identical, because in the new {\tt meb} model only the unique values of ${\tt w}$ are used. Thus, if two or
more elements of ${\tt w}$ are \emph{identical}, then they refer to the \emph{same} element in the covariate $\bm{x}$, which is not desired here.
Next, we define the {\tt meb} model formula. The model contains two hyperparameters:
\begin{itemize}
\item {\tt beta} corresponds to $\beta_x$, the slope coefficient of the error-prone covariate $\bm{x}$, with a Gaussian prior.
\item {\tt prec.u} is the error precision $\tau_u$ with gamma prior.
\end{itemize}

The prior settings are defined in the different entries of the list {\tt hyper}. The option {\tt fixed} specifies whether the corresponding quantity should be estimated or fixed at the {\tt initial} value. The field {\tt param} captures the prior parameters of the corresponding prior distribution. A Gaussian prior distribution is the default for ${\tt beta}$, while a gamma distribution is used for ${\tt prec.u}$ (again defined as log-gamma distribution for the log-precision).\\
The model contains as additional fixed effect the degree of defoliation {\tt z}, plus an additional i.i.d.\ random effects term per individual to account for unspecified heterogeneity, specified in {\tt f(individual,...)},  which extends the GLM to a GLMM:

\begin{Schunk}
\begin{Sinput}
 library(INLA)
 formula <- y ~  f(w, model="meb", hyper = list(
      	beta = list(
        		param = prior.beta,
        		fixed = FALSE
      	),
      	prec.u = list(
        		param = prior.prec.u,
        		initial = log(prec.u),
        		fixed = FALSE
      	)
    )) +
    z +
    f(individual, model = "iid", values = 1:n, hyper = list(prec = list(
  	    initial = log(prec.tau), 
  	    param = prior.tau
  	)
       )
    ) 
\end{Sinput}
\end{Schunk}

The call of the {\tt inla} function includes the specification of the family, which is Poisson here and thus includes no additional hyperparameters. The prior distributions for the intercept $\beta_0$ and the slope $\beta_z$ are specified in the {\tt control.fixed} option.

\begin{Schunk}
\begin{Sinput}
 r <- inla(formula, data = data.frame(y, w, z, individual),
           family = "poisson",
           control.fixed = list(
             	mean.intercept = prior.beta[1], 
             	prec.intercept = prior.beta[2],
             	mean = prior.beta[1],
             	prec = prior.beta[2]),
      )
 r <- inla.hyperpar(r)
 summary(r)
\end{Sinput}
\end{Schunk}

\vspace{10mm}
\nin{\bf Analysis with the {\tt copy} feature}\\[-4mm]

\nin
As described in the main text, as an alternative to the use of the new {\tt meb} model, the same results can be obtained by employing the {\tt copy} feature in INLA. The approach is similar to the one taken in Section~\ref{FramApp}. Recall though that in case of Berkson measurement error, the use of the {\tt copy} feature does not add to the generality of the model and is presented here only for completeness. \\

\nin
The object {\tt data} now contains an additional fourth column:
$$\bm{y} \, \quad \bm{w} \,\quad \bm{z} \,\quad  {\text{\bf sh}}$$

Column {\text{\bf sh}} contains the values sh$_1$, $\ldots$, sh$_n$, where sh$_i$ is the index of the shadehouse of seedling $i$. Note that the $n=60$ seedlings are distributed over $s=15$ shadehouses (1 $\leq$ sh$_i$ $\leq 15$), whereas always five shadehouses belong to the same light condition (dark, middle, light). There are thus 15 different correct light intensities ($\bm{x}$, one value per shadehouse), but only 3 different target light intensities ($\bm{w}$, one value per light condition).
As the error model in this example is Berkson, the joint model simplifies to two equations and the response matrix has only two columns. The model can be represented as 

\begin{equation}\label{jointModelB}
\underbrace{\begin{bmatrix}
y_{1} & \na   \\
\vdots &   \vdots \\
\vdots &   \vdots \\
y_{n} &   \na \\
\na & -w_1 \\
\vdots & \vdots   \\
\na & -w_s  \\
\end{bmatrix}}_{ {\tt Y}}
=  \beta_0   \underbrace{\begin{bmatrix} 1 \\  \vdots \\  \vdots\\ 1 \\ \na \\ \vdots   \\ \na \end{bmatrix}}_{ {\tt beta.0}}
+ \beta_x   \underbrace{\begin{bmatrix} \text{sh}_1 \\ \vdots \\ \vdots \\ \text{sh}_n \\ \na \\ \vdots   \\ \na \end{bmatrix} }_{ {\tt beta.x}}
+ \underbrace{\begin{bmatrix} \na \\ \vdots \\ \vdots  \\ \na \\ -1 \\ \vdots \\ -s \end{bmatrix} }_{ {\tt idx.x}}
+ \beta_z   \underbrace{\begin{bmatrix} z_1 \\ \vdots \\ \vdots \\ z_n  \\ \na \\ \vdots   \\ \na \end{bmatrix}}_{{\tt beta.z}}  
+  \underbrace{\begin{bmatrix} 1 \\ \vdots \\ \vdots \\ n  \\ \na \\ \vdots   \\ \na \end{bmatrix}}_{{\tt gamma}}   \ .
\end{equation}

Terms below the brackets correspond to the names in the R-code.\\
Let us start again with prior specification process in accordance to the main text:
\begin{Schunk}
\begin{Sinput}
 data <- read.table("shading_data4supp.txt", header=T)
 attach(data)
 w.red <- aggregate(w, by = list(sh), FUN = mean)[,2] 
 n <- 60 	# number of seedlings
 s <- 15 	# number of shadehouses
\end{Sinput}
\end{Schunk}

\begin{Schunk}
\begin{Sinput}
 prior.beta <- c(0,0.01)
 prior.tau <- c(1,0.005)  
 prior.prec.u <- c(1,0.02)   
\end{Sinput}
\end{Schunk}
\begin{Schunk}
\begin{Sinput}
 # initial values (mean of prior)
 prec.tau <- 1/0.005 
 prec.u <- 1/0.02        
\end{Sinput}
\end{Schunk}

The {\tt aggregate} command in the second line aggregates the vector $\bm{w}$ of length $n=60$ into the $15$ (one per shadehouse) unique light values.\\
Next, the response matrix {\tt Y} and the data vectors are filled according to the naming of Equation~(\ref{jointModelB}):

\begin{Schunk}
\begin{Sinput}
 Y <- matrix(NA, n+s, 2)
 Y[1:n, 1] <- y
 Y[n+(1:s), 2] <- -w.red
 beta.0 <- c(rep(1, n), rep(NA, s))
 beta.x <- c(sh, rep(NA, s))
 idx.x <- c(rep(NA, n), 1:s)
 weight.x <- c(rep(NA, n), -rep(1, s))
 beta.z <- c(z, rep(NA, s))
 gamma <- c(1:n, rep(NA, s)) 
 data.joint <- data.frame(Y, beta.0, beta.x, idx.x, weight.x, beta.z, gamma)
\end{Sinput}
\end{Schunk}

The definition of the INLA formula is almost analogous to the one in Section~\ref{FramApp}. The main difference is the additional i.i.d.\ random effects term per individual $\gamma_{ijk}$, specified in {\tt f(gamma,...)},  which extends the GLM to a GLMM:
\begin{Schunk}
\begin{Sinput}
 library(INLA)
 formula <- Y ~  beta.0 - 1 +
    f(beta.x, copy = "idx.x", 
  	hyper = list(beta = list(param = prior.beta, fixed = FALSE))) +
    f(idx.x, weight.x, model = "iid", values = 1:s,
  	hyper = list(prec = list(initial = -15, fixed = TRUE))) +
    beta.z +
    f(gamma, model = "iid", values = 1:n,
  	hyper = list(prec = list(initial = log(prec.tau), param = prior.tau))) 
\end{Sinput}
\end{Schunk}
As in Section~\ref{FramApp} we have to explicitly remove the common intercept using {\tt -1}.
The call of the INLA function is as well in analogy to Section~\ref{FramApp}, but there are only two likelihoods involved here: the Poisson likelihood for the regression model and the Gaussian likelihood for the error model. The former has no additional hyperparameters, while in the latter the error precision $\tau_u$ needs specification.
\begin{Schunk}
\begin{Sinput}
 r <- inla(formula, data = data.joint,
  	family = c("poisson", "gaussian"),
  	control.family = list(
             	list(hyper = list()),
             	list(hyper = list(
  			    prec = list(
  		            	initial=log(prec.u),
  			    	param = prior.prec.u,
  			    	fixed = FALSE)))),
  	control.fixed = list(
             	mean.intercept = prior.beta[1],
             	prec.intercept = prior.beta[2],
             	mean = prior.beta[1],
             	prec = prior.beta[2])
       )
 r <- inla.hyperpar(r)
 summary(r)
\end{Sinput}
\end{Schunk}

\eject

\section{Supplements to Sections 5.2 and 5.3 in the main text}

\subsection{Full conditionals for the MCMC sampler of Section 5.2}

Let all variables be defined as in Section 5.2 of the main text, see also 
the beginning of Section~\ref{FramApp} in this Supplementary Material for a
compact review.

The full conditionals for the unknowns in the regression model are given as follows:\\
For $\bm{\beta}=(\beta_0,\beta_x,\beta_z)^\top$ we have
\begin{eqnarray*}
\bm\beta \,|\, rest &\propto& \pi (\bm{y} \,|\, \bm{x},\bm{z}) \cdot \pi(\bm\beta)\\
	&\propto& \exp\left(\sum_{i=1}^n y_i \eta_i - \sum_{i=1}^n \log(1+e^{\eta_i}) -\frac{\tau_{\beta}}{2}  \bm\beta^\top\bm\beta  \right) \ .
\end{eqnarray*}
The $\bm\alpha=(\alpha_0,\alpha_z)^\top$ coefficients can be sampled from a Gaussian distribution.
Let $\mathbf{D}$ be the matrix with rows $D_i^\top := (1 \;\; {z}_i^\top)$, $\mathbf{R}=\begin{pmatrix}
	  \tau_{\alpha_0} & 0\\
	   0&\tau_{\alpha_z}
	\end{pmatrix}$, and $\bm{\mu} = (\mu_{\alpha_0}, \mu_{\alpha_z})^\top$. Then 
\begin{eqnarray*}
{\bm\alpha} \, | \, rest &\propto& \pi({\bm{x}}\,|\,rest)\cdot \pi({\bm{\alpha}}) \\
	&\propto& \exp\left( -\frac{\tau_{x}}{2} ({\bm{x}}-\mathbf{D}{\bm\alpha})^\top({\bm{x}}-\mathbf{D}{\bm\alpha})-
	\frac{1}{2}\left(\bm{\alpha} - \bm{\mu}\right)^\top
	\mathbf{R}
	\left(\bm{\alpha} - \bm{\mu}\right)
	 \right)\\
	&\sim& \mathcal{N}\left((\tau_x\mathbf{D}^\top \mathbf{D}+\mathbf{R})^{-1}\left(\tau_x\mathbf{D}^\top{\bm{x}} + \mathbf{R}\cdot \bm{\mu}\right), \tau_{x}\mathbf{D}^\top \mathbf{D}+
	 \mathbf{R} \right) ,
\end{eqnarray*}
where the second argument in the last expression again denotes the precision matrix. To sample from the distribution of the latent variable $\bm{x}$, full conditionals for $x_i$ are needed:
\begin{eqnarray*}
x_i \, | \, rest &\propto& \pi (y_i \,|\, x_i,z_i)  \cdot  \pi(w_{1i} \,|\, x_i) \cdot  \pi(w_{2i} \,|\, x_i)\cdot \pi(x_i \, | \, z_i) \\
	&\propto&  \exp\left(y_i \eta_i - \log(1+e^{\eta_i}) - \frac{\tau_u}{2}\left((w_{1i}-x_i)^2 + (w_{2i}-x_i)^2\right)  - \frac{\tau_{x}}{2}(x_i - \alpha_0 - \alpha_z z_i)^2 \right)  \ .
\end{eqnarray*}
Finally, the precisions can be sampled from gamma distributions
\begin{eqnarray*}
\tau_{x} \, | \, rest &\propto& \pi({\bm{x}} \, | \, \bm{z})  \cdot \pi(\tau_{x})  \\
&\sim& \text{G}\left(a_x +\frac{n}{2},b_x + \frac12 ({\bm{x}}-\mathbf{D}{\bm\alpha})^\top({\bm{x}}-\mathbf{D}{\bm\alpha})\right)  \ ,
\end{eqnarray*}
and
\begin{eqnarray*}
\tau_{u} \, | \, rest &\propto& \pi({\bm{w}} \, | \, \bm{x})  \cdot \pi(\tau_{u})  \\
&\sim& \text{G}\left(a_u + n,b_u + \frac12 (\bm{w}_1-\bm{x})^\top(\bm{w}_1-\bm{x})+ \frac12 (\bm{w}_2-\bm{x})^\top(\bm{w}_2-\bm{x})\right)  \ .
\end{eqnarray*}

\eject

\subsection{MCMC and INLA posterior marginals}
\begin{figure}[h!]
\centering
\mbox{\includegraphics[width=14cm]{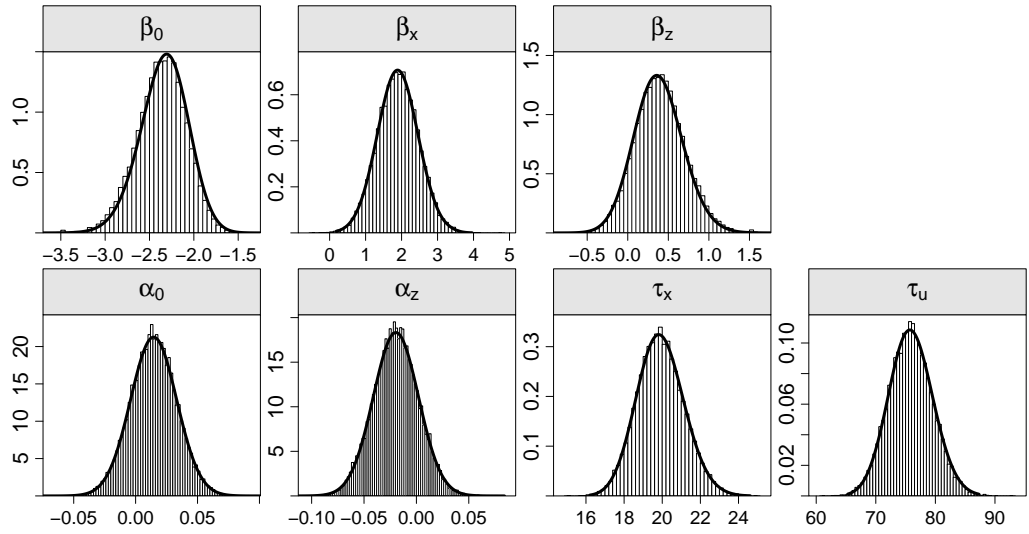}}
\caption{\label{FramHist}Comparison of the MCMC samples (histograms) with the INLA posterior marginals (lines) for the Framingham data (Section 5.2).}
\end{figure}

\begin{figure}[h!]
\centering
\mbox{\includegraphics[width=14cm]{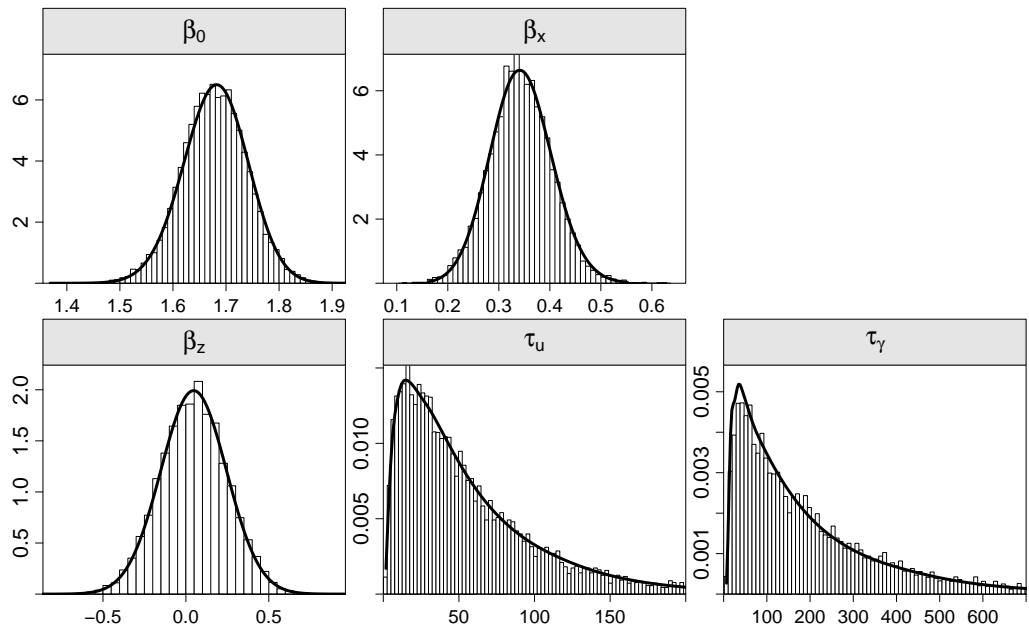}}
\caption{\label{SeedlingHist}Comparison of the MCMC samples (histograms) with the INLA posterior marginals (lines) for the seedling growth data (Section 5.3).}
\end{figure}

\end{document}